\documentclass[fleqn,usenatbib]{mnras}
\usepackage{balance}
\usepackage{newtxtext,newtxmath}
\usepackage{float}
\usepackage{graphicx}
\usepackage[T1]{fontenc}
\usepackage{ae,aecompl}

\usepackage{graphicx}	
\usepackage{amsmath}	
\usepackage[squaren]{SIunits}	
\usepackage{cuted}      
\usepackage[normalem]{ulem}
\usepackage{longtable}
\usepackage{dblfloatfix}   

\makeatletter					
\graphicspath{{./figures/}}		

\title{SCATTER Common Envelope Formalism for Triples}

\author[R. Di Stefano, A. Khwaja \& C. Kobayashi]{%
Rosanne Di~Stefano$^{1}$,
Amaan Khwaja$^{1}$,
and Chiaki Kobayashi$^{2}$\thanks{E-mail: \href{mailto:c.kobayashi@herts.ac.uk}{c.kobayashi@herts.ac.uk}}\\
$^{1}$Harvard–Smithsonian Center for Astrophysics, 60 Garden Street, Cambridge, MA 02138, USA\\
$^{2}$Centre for Astrophysics Research, Department of Physics, Astronomy and Mathematics, 
University of Hertfordshire, Hatfield, AL10 9AB, UK
}

\date{Accepted 21 January 2026. Received 21 January 2026; in original form 5 Nov 2025}

\pubyear{2025}

\begin{document}
\label{firstpage}
\pagerange{\pageref{firstpage}--\pageref{lastpage}}
\maketitle

\begin{abstract}
Many stars are components of triple-star systems, or of higher-order multiples. In such systems mass transfer is common, and when the transfer is dynamically unstable, a common envelope forms. As such, it is important to be able to compute the post--common-envelope orbital separations among the various stars comprising the system, and to determine whether the common envelope induces mergers or makes later mergers inevitable. In this paper we compute the results of common-envelope evolution for triples. We employ the SCATTER formalism, a new approach to the computation of post--common-envelope separations. This work has applications to gravitational-wave mergers, Type~Ia supernovae, and a broad range of other highly energetic phenomena.
\end{abstract}



\section{Introduction}
\label{sec:introduction}
A large fraction of massive stars are born in triples \citep{moe2017mind, duchene2013stellar, raghavan2010survey, TokovininTripleOrigin,tokovinin2014binaries}. Stars in many triples become close enough to each other that the system will pass through at least one stage of mass transfer or mass loss. Of particular interest are scenarios leading to the merger of two or more of a triple's stars. These scenarios generally require that the triple pass through at least one phase during which its components spiral closer to each other within a gaseous envelope that has been stripped from one of the companion stars. The envelope is called the {\sl common envelope} (CE).

The CE phase of higher-order multiples is important because each CE phase can lead to mergers and/or to smaller orbits. Additionally, subsequent evolution of CE end states can lead to further mass transfer, mergers, and/or additional CE episodes. The CE process and its end states yield high-mass objects, alone or in multiple systems, that can eventually either exchange mass or merge. Whether mergers occur within the CE or afterwards, they can create extreme luminosities in the electromagnetic and/or gravitational-wave regimes. Some mergers produce Type~Ia supernovae (SNe Ia), or accretion-induced collapse to an neutron star (NS) or black hole (BH). Triple-star CEs are expected to be common because massive stars have a high probability of starting as triple systems \citep{moe2017mind, Tokovinin_2018_MSC}, and triples may be needed to provide pathways to certain intriguing end states, such as binary BH mergers \citep{stegmann_2025} and massive BHs\footnote{This is because triples provide additional channels for interaction, beyond those available in binaries.}. Furthermore, triples and higher-order multiples are also formed through dynamical interactions in dense environments \citep{rasio1994binary,van2007formation, perets2012triple, leigh2011analytic, antonini2016black, martinez2020black,offner2022origin}.

\subsection{The CE in Binary Evolution}

Common envelopes play important roles in binary evolution. The CE was first introduced as a way to form cataclysmic variables \citep{pac76}, binaries in which a white dwarf (WD) accretes mass from a close low-mass companion. Despite decades of study, there are still many open questions about CEs. Some present-day studies \citep{passy2011simulating,ricker2012amr,ohlmann2015hydrodynamic,trani2022revisiting}
focus on the complex task of simulating the evolution of individual systems (see \cite{ivanova2020common}, Ch.~4, and \cite{ropke2023simulations} for reviews). The most frequent use of the CE is in the context of population-synthesis studies. Population synthesis starts with a population of stellar systems and evolves each system to determine the number and characteristics of interesting end states, such as gravitational-wave mergers or SNe~Ia. The evolutions are not the kind of detailed calculations one would do to predict the physical characteristics of each system over time. They are designed, however, to capture the significant features of stellar and binary evolution in a parameterized way, by using input from both theory and observations. While the end states predicted by the calculations may not be reliable guides to the characteristics of a particular individual system, the hope is that, on average, the ensemble of outcomes will represent realistic predictions for real populations of stars. Because a large fraction of ``interesting'' end-state systems pass through one or more intermediate states with a CE, population-synthesis calculations must be designed to map the state at the beginning of the CE to a state that has a good chance of realistically representing the final state of that or of a similar system. It has been noted that, both for the production of Type Ia supernovae through the mergers of WDs and for the production of compact-object mergers that involve BHs or NSs, the CE evolution plays a key role \citep{Belcz_DD_2009ApJ...699.2026R, Belcz_BH_2021A&A...651A.100O}.

In one set of CE formulations (the ``$\alpha$'' formulation), the basic underlying principle is conservation of energy \citep{van1976late,webbink1984,livio1988common}. The energy needed to disperse the CE is provided at the expense of gravitational binding energy. While conservation of energy is a powerful principle, the nature and relative contributions of the sources and sinks of energy are many and varied, including recombination/ionization energy and shocks.
An alternative approach is found in the ``$\gamma$'' formalism, which applies conservation of angular momentum to the binary system as a whole \citep{nelemans2000reconstructing,nelemans2005reconstructing}.

In both the $\alpha$ and $\gamma$ formalisms, a single value of $\alpha$ or $\gamma$ is typically used for each population-synthesis simulation. Different simulations are then carried out with different values of the CE parameters to test for the parameter dependence of the results. Note, however, that a small value of $\alpha$ may be ideal for some types of binaries, but not for others. Larger values of $\alpha$ will lead to correct results for other binaries. It is therefore unclear what we learn from comparisons among simulations that each use a single parameter value.

Here we apply the SCATTER formalism \citep{DiStefano_2023}, which is also based on the conservation of angular momentum. It allows {\sl each stellar component of the system} to exchange angular momentum with the envelope, whether the system is a binary, triple, or higher-order multiple. Like the other formalisms, it maps an initial, pre-CE state to a final, post-CE state via an analytic formula. It includes, however, parameters that can be modeled as functions of mass ratios that vary from system to system. The choice of SCATTER parameter functions for binaries has been calibrated based on a nearly complete set of post--common-envelope binaries containing at least one WD \citep{kruckow2021catalog}. We therefore know that the version of SCATTER presented here is capable of describing a wide range of WD-containing post-CE systems.

The formalism provides enough freedom to incorporate data appropriate to any set of binaries or triples, including those that lead to BH--BH, NS--NS, or BH--NS mergers. We do not know whether conducting a study of the post-CE states of such systems would produce the same set of parameterized functions derived for WD-containing systems. Such a comprehensive study is not yet possible, given the relatively small samples presently available for these massive systems. We therefore use parameter values derived from systems of lower mass and plan to incorporate new parameter values if needed, as more data become available.

The goal of binary CE calculations is to determine the binary separation, $a(f)$, at the end of the CE phase. Since we start by knowing the initial separation, $a(0)$, the single quantity we seek to compute may be thought of as the ratio $a(f)/a(0)$. Angular-momentum formalisms can, in principle, have three independent equations for angular-momentum conservation -- one for each component. At present, though, we generally do not know enough about the initial state of the system to productively invoke all three spatial components of the angular momentum. Fortunately, a single equation for the conservation of the total angular momentum is adequate to solve for the single variable needed for binaries. In this paper we show how the simple binary SCATTER CE formalism can be naturally extended to triples.

\subsection{The CE for Higher-Order Multiples}

Adding a third body makes the problem more complex. There are now three potentially independent separations to consider: $a_{12}, a_{13},$ and $a_{23}$. In principle, if we choose to consider the triple as a whole, we only have a single equation for total angular momentum, which alone is not enough to allow us to map the initial and final values of each of these three separations. Furthermore, it may be difficult to observationally measure the values of these separations either before or after the CE, especially if the separations are comparable to each other and the system is potentially chaotic.
Even stable triples can exhibit a range of interesting behaviors linked to the exchange of angular momentum. For example, the motion of the outer star can alter the eccentricity and orientation of the inner binary (\citealt{Kozai1962}; \citealt{LIDOV1962719}; see \citealt{Naoz2013} for a review).

The key assumption made by SCATTER is that, during the CE phase, interactions between the stars and the CE are the primary ways that angular momentum is drawn from or given to the stars. As such, while triple-star dynamics may be active in forming the pre-CE configuration, they do not govern the evolution of the CE itself. If no mergers occur during the CE phase, such dynamics may again become active after the CE has ceased. Alternatively, if the CE yields a system with close separations, angular-momentum loss due to the emission of gravitational radiation may become dominant post-CE. It is important to emphasize, however, that during the CE it is envelope interaction that acts as the dominant mechanism in transforming the pre-CE triple into its immediate post-CE state.

Furthermore, triples that respect a spatial hierarchy can be dynamically stable \citep{Szebehely1967,harrington1968dynamical,eggleton1995empirical}. Consider an inner binary composed of Stars~1 and 2, with masses $M_1$ and $M_2$, respectively. We denote their orbital separation by $a_{\rm in}$. If Star~3, with mass $M_3$, orbits the center of mass of the inner binary in a wide enough orbit, with semimajor axis $a_{\mathrm{out}}$, the system can be considered to be a hierarchical triple that is dynamically stable.
Furthermore, for hierarchical triples in a CE, the inner and outer orbits can be considered separately.
This is because the transfer of angular momentum to the envelope is more significant than transfers due to the direct interactions of the stars. Thus, rather than trying to solve for three unique orbital separations solely from the triple's total angular momentum, we instead subdivide the triple into separate binaries to which we can then apply SCATTER to solve for the change in orbital separation.

In Section \ref{sec:2} we provide the equations for binaries derived from the SCATTER formalism, and outline the elements we will use to extend SCATTER to triples. In Section \ref{sec:3} we discuss some general features of the extension of SCATTER to the case of triples. Sections~\ref{sec:4} and \ref{sec:5} are devoted to hierarchical triples. In Section \ref{sec:4} we present the case in which the star in the outer orbit fills its Roche Lobe (RL). In Section \ref{sec:5} we explicitly consider the case in which a star in the inner binary fills its RL. Section~\ref{sec:6} discusses the more difficult case in which the binary separations do not appear to form a natural hierarchy because the inter-star separations are all comparable in value. In Section \ref{section7} we study the robustness of the SCATTER calculations with respect to (1)~variations in sets of input parameters, and (2)~variations in the parameterized functions of the formalism. We devote Section \ref{sec:8} to our conclusions and to a discussion of future uses of SCATTER and possible directions for further development of the SCATTER formalism.

\section{The SCATTER Formalism for Binaries}{\label{sec:2}}

\subsection{Overview of SCATTER} 

In the SCATTER formalism we start with an $N$-body stellar system in a pre-CE state defined by the masses of the components and the distances between them. SCATTER maps this initial configuration into a post-CE configuration. If there are mergers during the CE, the number of system components may be reduced. The formalism determines if this is likely to be the case. It can also provide estimates of the inter-star separations for those stars that survive or are formed during the CE. 

The underlying assumption of SCATTER is that, during the CE, changes in the configuration of stars occur through interactions between each star and the matter in the CE. Thus, changes in the $N$-body system's angular momentum are caused by the interaction of each body with the CE. Changes in inter-star distances are associated with the changes in angular momentum. 
Here, we summarize the results of the calculations for binaries. Details can be found in \citet{DiStefano_2023}.

\subsection{SCATTER for Binaries}

We start with a binary containing masses $M_i(0)$ and $M_j(0)$, with semimajor axis $a(0)$. The total mass is $M_{\mathrm{tot}}(0)=M_i(0)+M_j(0)$. We then assume CE evolution occurs when Star~$i$ fills its RL. At this time the CE-donating star has a core mass equal to $M_i^c$ and an envelope mass equal to $M_i^{\mathrm{env}}$. We consider the initial state of the system to be that in which the envelope has 
just been released but still surrounds Star~$i$. Thus, at the very start of the CE, the mass of Star~$i$ is $M_i(0)=M_i^c+M_i^{\mathrm{env}}$. The core, $M_i^c$, is then considered as one component of a binary consisting of itself and the companion, Star~$j$. The masses $M_j$ and $M_i^c$ spiral toward each other within the envelope of mass $M_i^{\mathrm{env}}$.

We use the expression $q_{a,b}$ to denote the ratio $(M_a/M_b)$; thus, $q_{c,j}=M_i^c/M_j$. Each component of the inspiralling binary (i.e., Star~$j$ and the core of Star~$i$) interacts with the CE. 
The post-CE binary differs from the pre-CE binary both because the envelope has been lost and because each of the inspiralling components has transferred angular momentum to the CE. Depending on the relative contributions of mass loss and angular-momentum loss, the post-CE binary may be either larger or smaller than the pre-CE binary.

In the simplest case, the mass that the inspiralling components interact with is simply the mass of the envelope lost by the RL-filling star. There are cases in which the mass that serves as the source or sink of angular momentum for the binary components, $M^{\mathrm{interact}}$, is smaller or larger than the envelope mass. To begin, we set $M^{\mathrm{interact}}=M_i^{\mathrm{env}}$. We divide $M^{\mathrm{interact}}$ into two mathematically distinct parts, each receiving angular momentum through direct or indirect interactions with one of the inspiralling components. We make no assumptions about the location of the parts of the envelope most influenced by each inspiralling component. In fact, the only assumption we make about the CE is that it can act as a sink or source of angular momentum. 
The fraction of the envelope mass receiving angular momentum primarily from the inspiralling core is 
\begin{equation}
    {\cal Q}_{\delta}(q_{c,j})= \frac{f(q_{c,j})^\delta}{f(q_{c,j})^\delta+f(q_{j,c})^\delta}.
\end{equation}
The function $f(q)$ is given from Eggleton 1983 as
\begin{equation}
    f(q)=\frac{0.49q^{2/3}}{0.6q^{2/3}+\ln\left(1+q^{1/3}\right)}.
\end{equation}
Thus, the amount of mass interacting with Star~$c$ is $\bigl[ M^{\mathrm{interact}} \times {\cal Q}_\delta (c,j) \bigr]$. The fraction of the envelope mass receiving angular momentum primarily from the inspiralling Star~$j$ is of the same form, with $q_{j,c}=1/q_{c,j}$ replacing $q_{c,j}$. Note that ${\cal Q}_{\delta}(q_{c,j}) + {\cal Q}_{\delta}(q_{j,c}) = 1$. 
In principle, the value of the parameter $\delta$ is free. In practice, we will use a fixed value of $\delta$, $\delta=3$, for reasons described at the conclusion to this subsection. 
To simplify notation, we will refer to ${\cal Q}_{\delta}(q)$ 
as ${\cal Q}(q)$.

In the case we consider we have $M_i(f)=M_i^c$ and $M_{\mathrm{tot}}(f)=M_i(f)+ M_j(f)$, with $M_j$ constant. Imposing the condition that angular momentum transferred to the envelope is equal to the angular momentum lost by the binary, we derive an equation for the ratio $a(f)/a(0)$, where $a(0)$ is the orbital radius at the time the CE is released, and $a(f)$ is the post-CE orbital radius.
The form of the ratio $a(f)/a(0)$ is simplified by introducing the following function:
\begin{equation}
    {\cal F}(q_{c,j})= \frac{\eta_c}{q_{c,j}} {\cal Q}(q_{c,j}) +
    \frac{\eta_j}{q_{j,c}} {\cal Q}(q_{j,c}) = {\cal F}(q_{j,c}), 
    \label{eq:bigF}
\end{equation}
and
\begin{equation}
\begin{split}
\frac{a(f)}{a(0)} = 
\left(\frac{M_{\mathrm{tot}}(f)}{M_{\mathrm{tot}}(0)} \right)
\left(\frac{M_{i}(0)}{M_i(f)} \right)^2
\left(\frac{M_j(0)}{M_j(f)} \right)^2 \\
\times \exp\left[-\frac{2\, M^{\mathrm{interact}}}{M_{\mathrm{tot}}(f)} \, {\cal F}(q_{c,j}) \right].
\end{split}
\label{eq:afin_ratio}
\end{equation}
We have introduced the mass ratios $q_{a,b}$ in the function ${\cal F}$. $M_a$ and $M_b$ represent the masses that are spiralling toward each other within the CE. In the case considered here, $M_a=M_i^c$ and $M_b=M_j$. Note that, since ${\cal F}$ is symmetrical with respect to the interchange of $q$ and $1/q$, the choices of $M_a$ and $M_b$ could instead have been made to have $M_b=M_i$ and $M_a=M_j^c$.

The functional form of ${\cal F}$ introduces two new parameters, $\eta_c$ and $\eta_j$. In the calculations we describe here we will, as in \citet{DiStefano_2023}, use a single value: $\eta = \eta_c = \eta_j$. The value of $\eta$ is a measure of the efficiency with which angular momentum is transferred to the envelope from the individual stars. Larger values of $\eta$ are associated with more orbital shrinkage because the efficiency of angular-momentum transfer is lower. 

In order to find appropriate values of $\eta$,
\citet{DiStefano_2023} used the equation for $a(f)/a(0)$ to map the pre-CE state to the post-CE state of 112 post-CE binaries from \citet{kruckow2021catalog}. This allowed a functional form for $\eta$ to be derived as
\begin{equation}
\log_{10}[\eta] = -A\, \log_{10}\!\left[\frac{M^{\mathrm{interact}}}{M_{\mathrm{tot}}(f)}\right] + B.
\end{equation}
$M_b(f)$ is the final mass of the inspiralling binary, $M_1^c+M_j$. For the simple case of an isolated binary, $M^{\mathrm{interact}} = M^{\mathrm{env}}$.
Values of $A$ and $B$ are computed from the fit to the data obtained from the post-CE binaries. We fit the values of $A$ and $B$ for post-CE binaries of different types and also for binaries that experienced different amounts of shrinkage during the CE. Typical values are $A = 0.95$ and $B= 0.6$, which we take as the fiducial values in this paper. 

In \citet{DiStefano_2023} a range of $\delta$ values was considered. In order to avoid a large dependence on the value of $\eta$ needed to achieve a specific end state, we found $\delta = 3$ to be a good choice. This value is also associated with a clear physical interpretation, as $(a \times f(q))^3$ (where $a$ is the orbital separation) represents a volume. 

Note that, in the SCATTER approach, we make no assumptions about the fate of the CE once it is lost. Our entire focus is on angular-momentum transfer between the CE and each stellar component spiralling within it.

\subsection{Goals of the CE Calculations}

The questions we are most interested in answering are: what is the state of the system immediately after the CE? and, what is the eventual fate of the system?

There are uncertainties in each CE formalism, producing uncertainties in the final state of every individual system. The hope is, however, that when a formalism is applied to a large group of systems, the results will be statistically equivalent to what would have been derived if we could actually carry out the calculations in detail.

In formulations like the $\alpha$ formalism, a suite of simulations using different values of $\alpha$ are often carried out. By conducting calculations for different values of $\alpha$, the efficiency of energy transfer, from low to high values, astronomers aim to determine the uncertainty associated with our value of $\alpha$. As mentioned earlier, one potential downside of this approach is that while some systems may be appropriately modeled with low values of $\alpha$, others may require high values. Thus, by using a single value of $\alpha$ in a simulation, one is correctly treating a certain subset of the systems, but not other subsets. By changing $\alpha$, we are conducting calculations that are correct for a changing, but generally unidentifiable, subset of systems. As a result, the uncertainties in the results of population-synthesis calculations using either the $\alpha$ or $\gamma$ formalism are not well defined and are difficult to quantify.

In contrast, the SCATTER formalism introduces a variable $\eta$ whose value is a function of mass ratios involving the binary masses and the envelope mass. The free parameters in this function are fit by the data. Although there are still inherent uncertainties in the parameters, they are better tuned to different types of systems within each simulation. 

We note that the SCATTER formalism often predicts less shrinkage than the $\alpha$ or $\gamma$ formalism \citep{DiStefano_2023}. SCATTER even predicts expansion in cases for which mass loss is more important than angular-momentum loss. Nevertheless, the amount of shrinkage predicted by SCATTER is enough to produce both prompt and delayed mergers for a wide range of systems. 

In this paper, we turn to three-body systems, for which there are not enough data to recalibrate the parameters derived for binaries. We therefore take care to show that the orbital changes we predict are robust with respect to changes in the parameters; and that gradual changes in the properties of the triples produce smooth changes in the computational results. Finally, we note that we focus on general results: does a system merge within the CE? after the CE? not merge, but shrink enough to facilitate future interactions? or does it not change in a way that significantly affects its future evolution? 

Finally, we note that any CE formalism which predicts the ratio $a(f)/a(0)$ can also predict $a(f)$. If we are armed with the values of the radius and total mass of the RL-filling star, as well as the mass of its companion, then the RL-filling condition will provide the value of $a(0)$. Thus, the calculation of the ratio of final to initial orbital radii generally allows $a(f)$ to be computed. As we discuss in Section 3, there are also conditions on the initial orbit of a third star that can be used to help determine the post-CE fate of the triple.

\section{The CE in Triple Systems}{\label{sec:3}}

\subsection{Approach to Triples}

We start with three stars, Star~1, Star~2, and Star~3.   
The distances between the stars are $a_{12}$, $a_{13}$,
and $a_{23}$. 
The initial mass of the system is $M_{\mathrm{tot}}(0)=M_1(0) + M_2(0) + M_3(0)$. This corresponds to an initial state in which mass has not yet moved from the vicinity of the RL-filling star, even though the envelope, of mass $M^{\mathrm{env}}$, is no longer bound to it. The final mass of the RL-filling star is its core mass, $M^c$. It is this core that spirals in toward its companions during the CE. We consider only cases in which no part of the CE is accreted by a member of the system. This means that the final mass of the system is $M_{\mathrm{tot}}(f)=M_{\mathrm{tot}}(0)-M^{\mathrm{env}}$.

\subsection{Orbital Stability and Hierarchical Triples}

The three-body system, acting under only the force of gravity, does not have a closed-form solution. Under a wide range of parameters, we know that triple-star dynamics produces instabilities in which stars may merge or else be ejected from the system. 
Despite this, stability remains achievable in some hierarchical triples. These are triples that can be viewed as a combination of two binaries: (1)~an inner binary, with semimajor axis $a_{\mathrm{in}}$; and (2)~a third star of mass $M_3$ in a wider orbit.  
The distance between Star~3 and the center of mass of the inner orbit is $a_{\mathrm{out}}$.
 
Stability conditions have been derived, allowing us to identify stable triples.  
We use the stability criterion of \cite{Mardling1999} for prograde coplanar triples, which states that
\begin{equation}
    G=\frac{a_{\text{out}}}{a_{\text{in}}} \Big{|}_{\text{crit}}=2.8 \left[(1+q_{\text{out}})\frac{(1+e_{\text{out}})}{(1-e_{\text{out}})^{1/2}}\right]^{2/5},
    \label{eq:Mardling}
\end{equation}
where $a_{\mathrm{out}}$ and $a_{\mathrm{in}}$ are the semimajor axes of the outer binary and the inner binary, respectively. We use $q_{\mathrm {out}} =\frac{M_3(0)}{M_1(0)+M_2(0)}$. The eccentricity of the outer orbit is $e_{\mathrm{out}}$. Thus, to ensure stability, the value of $a_{\mathrm{out}}$ must be greater than $G\times a_{\mathrm{in}}$. Because the stability condition requires that $a_{\mathrm{out}}$ be significantly larger than $a_{\mathrm{in}}$, it is appropriate to refer to systems satisfying the stability condition as hierarchical. When the stability condition is satisfied, calculations that treat the inner and outer binaries separately are likely to produce reasonable approximations to the results achieved by real triples. In this paper we set the eccentricity to zero. 
The form of Equation 6 assumes co-planarity, although not all systems undergoing a CE phase are co-planar.  It has been shown, however, that the form of the stability criterion is not a strong function of orbital inclination \citep{eggleton1995empirical}.

In order for there to be a CE, one of the stars must fill its RL. Since the RL is defined in terms of the gravitational force and rotation of a pair of stars, the RL-filling star is strongly affected by tides prior to RL filling. This means that the standard requirements for orbital stability may not be the correct ones to apply. Nevertheless, for the purposes of this paper we will employ the condition discussed above. 

When we are determining the stability of the pre-CE state, the quantities in Equation~\eqref{eq:Mardling} have their initial (pre-CE) values. It may happen that the CE does not lead to a merger, so that the post-CE state is also a triple. In such cases it is important to assess that triple's orbital stability in order to predict its future evolution. Thus, one may also employ Equation~\eqref{eq:Mardling} to the post-CE triple. However, when the components of the triple are close enough that general-relativistic considerations are important, Equation~\eqref{eq:Mardling} will not apply. In such cases, the time-to-merger predicted by general relativity may provide the most useful guide to whether there will be post-CE interactions or mergers. 

\subsection{The Roche Lobe}

The concept of the Roche lobe (RL) emerged from the study of close binaries. When a star fills its RL, it is poised to transfer mass to a co-rotating companion.  
The RL formalism is modified for triples \citep{dynamicalrl}. The effects of the modification include a moving $L_1$ point. Changes from the standard formalism become more pronounced when the separations between the stars in the triple are of similar sizes. The conditions for the equivalent of RL filling are not significantly affected, especially for hierarchical triples, and we will use them here.
The RL-filling star (i.e., the star that contributes its envelope) may be a member of either the inner or outer binary.

The fact that a star fills its RL does not necessarily imply that there will be a CE. A CE occurs when the process of mass transfer is unstable. If the effects of mass transfer from the donor cause the donor to expand faster than the RL can expand, the envelope of the RL-filling star can be stripped from it.  
This type of instability can occur when the donor star is a giant or subgiant, and can happen even for main-sequence donors if the mass ratio between the donor and its companion is high enough. 

\subsection{Modeling Triples for CE Calculations}

When the triple is hierarchical, the outer orbit is significantly larger than the inner orbit. Furthermore,
when there is a CE, interactions with the CE are more significant than pure dynamical interactions.  
For hierarchical triples, we therefore employ the approximation that the CE formalism can be applied separately to the inner binary and to the outer binary. As we will see, however, the two binaries share mass from the envelope, and the calculations and results reflect this. 

When the triple is not hierarchical, the problem is more complex. We can consider the instantaneous configuration of the triple as a triangle, with the three stars at the vertices, and three sides, $a_{12}$, $a_{13}$, and $a_{23}$. Thus, there are three 
separate separations to consider, all comparable in value, to within a factor of a few.  

\subsection{Post-CE States: Binaries}

While CE formalisms are not always accurate predictors of the post-CE physical parameters of individual systems, such formalisms aim to provide results that mirror reality for a broad ensemble of systems. What we can hope to do for a post-CE binary is to determine answers to the following questions. {\bf (1)}~Was there a merger between the components of the binary during the CE? {\bf (2)}~If so, what is the result of the merger? {\bf (3)}~If not, are the post-CE stars close enough to each other to interact during a future epoch? The common element in each case is that one of the two objects is the compact core of the star whose RL-filling triggered the CE. The second object during the CE may be either another compact object or else an extended star.

\begin{enumerate}
\item \textbf{Is there a merger during the CE?} 
A merger during the CE will occur if the inspiral time of the binary due to drag is shorter than the time needed to expel the envelope. For the case of an input–output style formalism such as SCATTER we select binaries in which the time to merger post CE is less than $(10^4$--$10^5)$~yr (typical dispersion time for CE). Thus we assume that a merger takes place before the CE has fully dispersed, even if no envelope material remains near the site of the merger. When one component of the binary is an extended star, then the merger criteria are related to the stellar radius. The criteria depend on the nature of the star, but mergers are generally expected when the distance of closest approach is less than a few stellar radii. To be specific, we will say that a merger occurs when the extended star would overfill its RL at closest approach. For a binary whose only components are BHs or NS, we say that a merger will occur during the CE if the final orbital separation yields a time to merger shorter than $10^4$ years. This value is a proxy for the lifetime of the CE; although our choice is somewhat arbitrary, the value can be tailored to suit specific types of binaries as needed.

\item \textbf{If there is a merger, what is the result?} 
If the two components are compact objects, gravitational and possibly electromagnetic energy and a variety of particles will be released upon merger. The result of a BH–BH merger is a BH, as is the result of a BH–NS merger. For a BH–WD or a NS–WD system, the WD will be tidally disrupted and a significant fraction of the mass will be accreted by the WD's companion. In general, the merger of two NSs will produce an object more massive than the maximum NS mass, potentially leading to an accretion-induced collapse (AIC) to a BH \citep{bernuzzi2020neutron}. The merger of two WDs can produce a SN Ia. Alternatively, WD mergers can lead to AIC to a NS, or simply to fast-rotating WDs or RCB stars \citep{yungelson2017merging, dan2014structure}. When one component is a main-sequence star or giant, mergers can yield fast-rotating giants, luminous red-nova–like transients, or (in WD/NS–giant cases) Thorne–Żytkow objects \citep{ablimit2022stellar}.
\item \textbf{If there is no prompt merger, will the post-CE binary interact later?} 
As above, the answers to this question depend on the natures of the two objects that remain in orbit with each other. For compact objects we need only ask if the time to gravitational-wave merger is smaller than a Hubble time. If so, they will merge, producing the end states described above, but at a later time. If one of the stars is extended, then we must ask whether it will fill its RL in a Hubble time, thus leading to a second epoch of mass transfer that, depending on the relative masses and state of evolution of the RL-filling star, could lead to either a second CE or else an epoch of stable mass transfer. If there is no future RL filling, mass may be transferred through winds, depending on the mass of the extended star and the distance between it and its compact companion.
\end{enumerate}

\subsection{Post-CE End States: Triples} 

The discussion above directly applies to binaries. The end results of mergers are single objects which may have high spins, and which may also have experienced a dramatic transformation upon merger, especially if the merger remnant exceeds a critical mass. A similar set of considerations comes into play when the CE occurs within a triple system.  
As above, one of the components is the core of the star that filled its RL, sparking the CE phase. The other two stars may consist of any combination of compact objects and/or extended stars. 
The final result can be one of the following.

\begin{itemize}
    \item A single merged object. This may be either a compact object or else an extended star whose core may be an ordinary nuclear-burning core, or else a BH or NS core. In any of these cases, the composite object is likely to display high rates of rotation.
    \item A binary. {\bf (a)}~The binary may consist of two compact objects; in this case, one of them is a merger remnant that is likely to have high spin. If the two compact components are close enough to each other, they may merge within a Hubble time, producing a single high-spin compact object, as mentioned above. 
    {\bf (b)}~The binary may instead contain an extended star with an ordinary nuclear-burning core, or else a BH or NS core. In this case predictions of future interactions must take into account the evolution of the extended star and its companion (which may be either a compact object or an extended star). There may be
a future epoch of mass transfer or even a CE.
    \item A triple will result in cases in which no merger occurs. If the stability limit is respected and the triple is dynamically stable, then the triple end state may be long lived. If the orbits are close enough that angular momentum is drained through the emission of gravitational radiation during the system's evolution, then the effects of general relativity must be taken into account to assess stability. In either the relativistic or non-relativistic case, the triple may be chaotic. In this case, the system's orbital evolution could lead to a triple-star merger or a double-star merger. Whether or not there is a merger, one of the three stars can be ejected at high velocity. If chaos reigns, then we can give only a probabilistic estimate of the outcomes \citep{stone2019statistical, toonen2022stellar}.
\end{itemize}

\smallskip

Note that we are considering that interval in the evolution during which the interactions of the CE with the multiple-star system dominate. There are cases, particularly when two components of the system are NSs or BHs, when gravitational effects can become more important, especially during close approaches. 
For example, one may speculate that depending on the natures and relative masses of the triple's components at the time that two of the stars merge, the merger could cause a significant recoil, a gravitational-wave ``kick'' with enough energy and momentum to potentially expel the remaining companion if the efficiency of energy and momentum transfer is high enough (see also \citealt{naoz_2025}).
This possibility should be included in calculations when the natures of the objects undergoing potential mergers are known to include any combination of NSs and BHs. The SCATTER formalism (or any of the other CE formalisms) can be employed up until the time when general-relativistic effects become as significant as the CE in changing orbital separations.

\subsection{The Functions $\eta$ and ${\cal Q}$}
{\label{sec:etainner}}
One of the most important aspects of the SCATTER formalism is the choice of the functional form of the angular-momentum-transfer efficiency, $\eta$. In \cite{DiStefano_2023}, $\eta$ was fit to known post-CE binaries. At present, these binaries have provided the only guide to the functional form of $\eta$. We note, however, that the collection and use of more data, even for WD-containing binaries, could influence $\eta$. We may find, for example, that we obtain better fits to the data by employing models in which the two terms in Equation~\eqref{eq:bigF} have different values of $\eta$. For binaries containing an NS or BH, the values of the parameters $A$ and $B$, used to compute $\eta$, are likely to be different from those for WDs.

Similarly, the functional form of $\eta$ in triples, in which one component is a star or stellar remnant and the other is a binary, is also likely to differ from the WD systems used to derive the values of $\eta$.  
For example, the gas dynamics and torques experienced by an inner binary embedded in the envelope of a tertiary may differ from the binary case. 
Ideally, for all types of binaries and triples, we would be able to use data on pre- and post-CE states to derive the functional form of $\eta$.  

There are two paths toward finding the most physically realistic functional form and parameterization of $\eta$. One is additional observational data. Eventually, the necessary data will be collected and analyzed. For example, GAIA is making it possible to identify a range of binaries in which one component is a compact object \citep{el2021million, shahaf2023triage,yamaguchi2024wide}. Thus, we can expect that in the coming years, enough data will exist to constrain $\eta$ for a range of binary types (e.g., those containing BHs).  

The second method is through comparison with simulations of individual systems. 
Although post-CE triples are generally not well constrained, several numerical studies have examined hierarchical triples in which the tertiary donates an envelope that engulfs an inner binary \citep{comerford2020estimating,glanz2021simulations,rosselli2024evolution}. A qualitative trend is that the outer orbit can shrink rapidly, while the inner binary can either (i) shrink and sometimes merge or (ii) expand and even become dissociated, depending on system parameters. This behavior points to $\eta$ values for triple systems that can differ from those inferred for binaries. Consider hierarchical triples in which the outer star contributes the CE. For the outer binary, one expects a broadly similar range of $\eta$ (the underlying physics of drag and envelope torques is similar), though possibly with slightly reduced angular-momentum transfer efficiency due to the presence of the inner pair \citep{glanz2021simulations}. The $\eta$ formalism can also predict expansion of the inner binary if mass loss from one or both of its components is allowed, or if some systems yield a negative value of $\eta$, for example if the orbits are retrograde. 

In fact, detailed computations of the evolution and fate of individual systems are complementary to SCATTER and other CE formalisms. For population synthesis it is, at present, computationally unrealistic to evolve each CE fully.
Despite this, information from detailed simulations can inform SCATTER and {\sl vice versa}. Robust trends in the relationship between the pre- and post-CE states, from the simulations, should be reproducible via SCATTER. Just as data from the sky can constrain the angular-momentum efficiency, so too can data from simulations. Even better, if simulations can track, for each mass element $dm$ in the simulation, the probability, $P_j$, that it exchanged angular momentum with Star~$j$,
the sum of the probabilities would allow us to compute $\eta$,
constituting a more fundamental way to calculate its value. Note, however, that observational data provide the results of real evolutions. Simulations at present must make a variety of physical assumptions, not all of which mimic nature.

As an example of how complexities in local physics affect the results of simulations we can
follow the results of \cite{rosselli2024evolution}. They find that the change in inner-binary separation depends sensitively on the dimensionless density-gradient parameter, which we denote $\epsilon_\rho$, of the surrounding flow: shallow gradients tend to harden (shrink) the inner binary, while steep gradients tend to soften (expand) it. Hence if we were to follow the results of such simulations there would be an inherent relationship between the density gradient and our $\eta$ parameter.

Thus further data and data analyses, along with more results from simulations, will help us better constrain the functional form and values of $\eta$ in complex systems. For this paper we choose to use the results from the observationally supported WD data \citep{DiStefano_2023}. We then explore the uncertainties by varying the values of the parameters $A$ and $B$ that were derived through fits from the 
data (Section \ref{section7}). 

SCATTER makes another important choice, which is the amount of matter in the CE that exchanges angular momentum with each component of the binary. This choice is made by designing a functional form for the variable $\cal Q$. The choice we have made is based on the RL formalism. The formalism can be useful for triples, although it is less likely to directly apply to non-hierarchical triples. Future work may suggest a different functional form. We note, however, that if the functional form used in a calculation leads to too much mass being assigned to interact with one star, and too little to interact with the other, the error can be compensated for via changes in the value of $\eta$ for each star. Thus, the use of data-based derivations of the functional form of $\eta$ will allow SCATTER calculations to yield realistic results, even if the functional form of $\cal Q$ is not optimal.

In deriving a form for ${\cal Q}$, we have assumed that all of the envelope's mass exchanges angular momentum with the system's stellar components. That is
$\sum {\cal Q}_j =1$, where the sum is over all of the stars in the system. If some mass leaves the system without exchanging (either donating or extracting) angular momentum, the sum is smaller than unity. This affects the final separations between the stars post-CE. For example, the loss of mass without the loss of angular momentum will lead to larger inter-star distances in the post-CE phase. 

As a final caveat, we note that, depending on the nature of the stars, the initial orbital separations may help to determine the value of ${\cal Q}$. There are many ways in which this dependence can express itself. An approach that is natural to the formalism is to consider the ratios of the initial orbital separations. If these always appear as a multiple of the appropriate mass ratio, we can employ the variables $Q_{i,j} = a_{i}/a_{j} \times q_i$\footnote{An alternative is to define $Q_{i,j} = a_{j}/a_{i} \times q_i$.}. The form of the equations shown in Section \ref{sec:2} remains the same, while the results of the calculations will reflect the introduction of separation dependence.  

\section{Hierarchical Triple: Outer Star Fills its RL}{\label{sec:4}}

The outer star is Star~3. Its initial mass is $M_3(0)=M_3^c + M_3^{\mathrm{env}}$, where $M_3^c$ is its core mass and $M_3^{\mathrm{env}}$ is its envelope mass. Its final mass is $M_3(f) = M_3^c=M_3(0)-M_3^{\mathrm{env}}$. Star~3 is in a wide orbit ($a_{\mathrm{out}}$) with the center of mass of the inner binary. The inner binary is defined by Stars~1 and~2. Its total binary mass is $M_{\mathrm{binary}} = M_1(0)+M_2(0)$. $M_{\mathrm{binary}}$ is constant, since $M_1=M_1(0)=M_1(f)$ and $M_2=M_2(0)=M_2(f)$. In this section we consider the case in which Star~3 fills its RL and mass transfer is dynamically unstable, leading to a CE.

\subsection{Orbital Changes} 

\subsubsection{Changes in the Outer Orbit}

The entire envelope mass, $M_3^{\mathrm{env}}$, interacts with the components ($M_3^c$ and $M_{\mathrm{binary}}$) of the outer orbit.
Thus, $M^{\mathrm{interact}}_{\mathrm{out}}$ for the outer binary is $M_3^{\mathrm{env}}$, and we get
\begin{equation}
\frac{a_{\mathrm{out}}(f)}{a_{\mathrm{out}}(0)}
=
\left(\frac{(M_1+M_2)+M_3^{c}}{(M_1+M_2)+(M_3^c+M_3^{\mathrm{env}})}\right) 
\left(\frac{(M_3^c+M_3^{\mathrm{env}})}{M_3^{c}}\right)^2 \times Y,
\label{eq:7}
\end{equation}
where
\begin{equation}
    Y = \exp\left[-\frac{2\, M_3^{\mathrm{env}}}{(M_1+M_2)+M_3^c} \, {\cal{F}}(M_3^c/M_{\mathrm{binary}})\right].
\end{equation}

\subsubsection{Changes in the Inner Orbit}

Neither component of the inner orbit changes its mass. Each component interacts, however, with mass that was once part of the envelope of Star~3. The fraction of $M_3^{\mathrm{env}}$ that interacts with the stars of the inner orbit is
\begin{equation}
    M^{\mathrm{interact}}_{\mathrm{in}} = M_3^{\mathrm{env}} \times {\cal Q}(M_{\mathrm{binary}}/M_3^c),
\end{equation}
and we get
\begin{equation}
\frac{a_{\mathrm{in}}(f)}{a_{\mathrm{in}}(0)} = \exp\left[-\frac{2\, M^{\mathrm{interact}}_{\mathrm{in}}}{M_{\mathrm{binary}}} \, {\cal{F}}(M_1/M_2)\right].
\label{eq:10}
\end{equation}

\subsection{Examples}

\subsubsection{Mapping the Pre-CE State to the Post-CE State}
We consider the case in which the inner binary is comprised of stars with masses $M_1(0) = 0.83\, M_\odot $ and $M_2(0) = 0.70\, M_\odot$. We need not make any assumption about the natures of the stars in the inner binary. Since, however, the outer star fills its RL, we need to specify its core mass and envelope mass at the time of RL filling. We will take $M_3(0)$ to be $5\, M_\odot$, large enough that in most cases it will lose mass on a dynamical time scale when it comes to fill its RL. We apply Equations~\eqref{eq:7} through \eqref{eq:10}, allowing the core mass of Star~3 to vary from $0.2\, M_\odot$ to $1.0\, M_\odot$; the latter is roughly equal to the maximum possible core mass for a $5\, M_\odot$ star.

\begin{figure}
\includegraphics[width=\columnwidth]{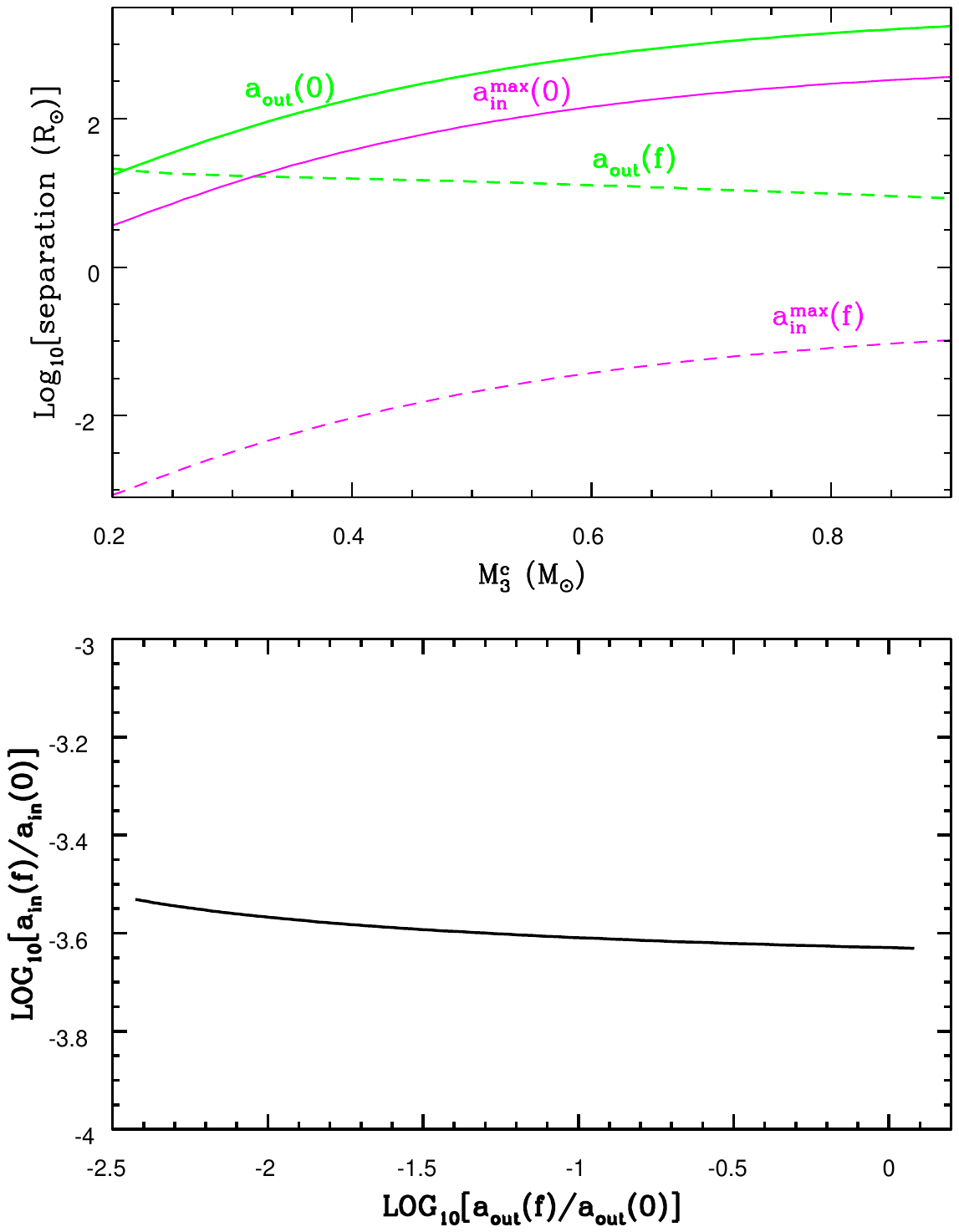}
\vspace{-0.4 true in} 
    \caption{{\bf Top panel:} Initial and final orbital separations versus $M_3^c$. The initial (post-CE) orbit of Star~3 has radius $a_3(0)$ ($a_3(f)$). Values of $a_3(f)$ range from tens of solar radii, for the smallest core masses, to just under $10\, R_\odot$ for the largest possible core mass. The star occupying the post-CE orbit is a WD, with radius too small to merge with other stars, unless the other stars are large. In this case, the other stars are Star~1 and Star~2, whose orbital radius (i.e., the radius of the inner orbit) shrinks by a factor as small as, or even smaller than, $0.0003$. Thus, the inner binary will merge. {\bf Bottom panel:} $\log_{10}\!\left[a_{\mathrm{in}}(f)/a_{\mathrm{in}}(0)\right]$ versus $\log_{10}\!\left[a_{\mathrm{out}}(f)/a_{\mathrm{out}}(0)\right]$ for varying tertiary core mass. }
    \label{fig:outer_fill}
\end{figure}

The results are shown in Figure~\ref{fig:outer_fill}. 
The uppermost (solid green) curve corresponds to the function $a_{\mathrm{out}}(0)$, the initial separation between Star~3 and the center of mass of the inner binary. Given the stellar mass, $M_3(0)$, each value of the core mass of Star~3 corresponds to a specific value of the stellar radius $R_3(0)$. This allows the orbital separation to be computed. Hence the separation is
\begin{equation}
    a_{\mathrm{out}}(0)=\frac{R_3(0)}{f(M_3(0)/M_{\mathrm{bin}}(0))},
\end{equation}
where $M_{\mathrm{bin}}(0)$ is the initial mass of the inner binary. The dashed green curve below, which is almost a straight line, is $a_{\mathrm{out}}(f)$, the final value of the separation. Larger values of the core mass of Star~3 are associated with more shrinkage, up to about two orders of magnitude for the largest core mass. Nevertheless, the smallest value of $a_{\mathrm{out}}(f)$ is roughly $10\, R_\odot$.  
Before discussing what this means for the future evolution of the triple, we consider the evolution of the inner binary.

The second (solid magenta) curve from the top shows $a_{\mathrm{in}}^{\max}(0)$, the maximum initial radius of the inner orbit consistent with three-body stability while also allowing the outer star to be RL filling. The dashed magenta line at the bottom of the panel shows the corresponding final inner-binary orbital radius, $a_{\mathrm{in}}^{\max}(f)$. We see that, for values of the third star's core mass as low as $0.2\, M_\odot$, the final inner orbital radius is $10^{-2.5} R_\odot$. With such a value, the components of the inner binary will merge within the CE, whether they are extended stars or compact objects. For example, even for the maximum value of $M_3^c$, the time to gravitational-wave merger is only $\sim$$17{,}000$~years, also indicative of a likely merger within the CE.

Thus, across all possible values of the physical parameter of Star~3's core mass, the result is a merger of the inner binary.\footnote{Note, however, that if the inner binary is able to lose mass, its shrinkage will be moderated; the inner binary could even expand.} Recall that we have considered the maximum possible value of $a_{\mathrm{in}}(0)$. Thus the selection of smaller initial values of the orbital radius will lead to smaller final values. 
This analysis shows that there are a variety of interesting post-CE states. In addition, for each possible trio of stars in this particular system, the character of the results is stable with respect to changes in the system parameters, in this case the core mass. This is a result of the stability discussed in Section \ref{section7}. 

\subsubsection{The Nature(s) of the Components, Possible Events, and Post-CE States}

The result of the merger of the components of the inner binary depends on the physical natures of the stars. If both are WDs, then they are CO WDs whose merger will lead to an object with likely mass above the Chandrasekhar mass. The merger is likely to lead to an SN~Ia. This explosion will take place within the CE, potentially producing spectral signatures of interactions with the circumbinary material. The third star will have a mass equal to that of its core at the time of RL filling. The orbital separation, $a_{\mathrm{out}}(f)$, will be at least $10\, R_\odot$ at the time of explosion and the core will be ejected with a velocity close in value to its orbital velocity.

If both of the components of the inner binary are extended stars, they are likely to both be on the main sequence. Their merger will also likely be a main-sequence star, but one of higher mass than the mass of either component. This star will be in an orbit with the core of Star~3, at a distance from $10\, R_\odot$ for a solar-mass core, to a few tens of $R_\odot$ for a He core. Interestingly enough, in this configuration, the merged main-sequence star will come to fill its RL as it evolves. If the core (i.e., the remnant of Star~3) has a relatively low mass, there will be a CE that envelopes both the evolving star and the core. The result may be the merger of two He stars or a He star and a CO WD. If, however, the core of the third star is more massive and thus becomes a CO WD, there could be stable mass transfer, possibly at a high enough rate to promote nuclear burning on the surface of the CO WD. This could produce an SN~Ia through the single-degenerate channel. 

If just one of the stars in the inner binary is a WD and the other is a main-sequence star, their merger could produce a giant that would unstably fill its RL with respect to the core of Star~3 even during the CE. The result would then likely be a merger of the core of the merged giant and the core of the original RL-filling star.

\subsubsection{Physical Implications}

\begin{figure}
\includegraphics[width=\columnwidth]{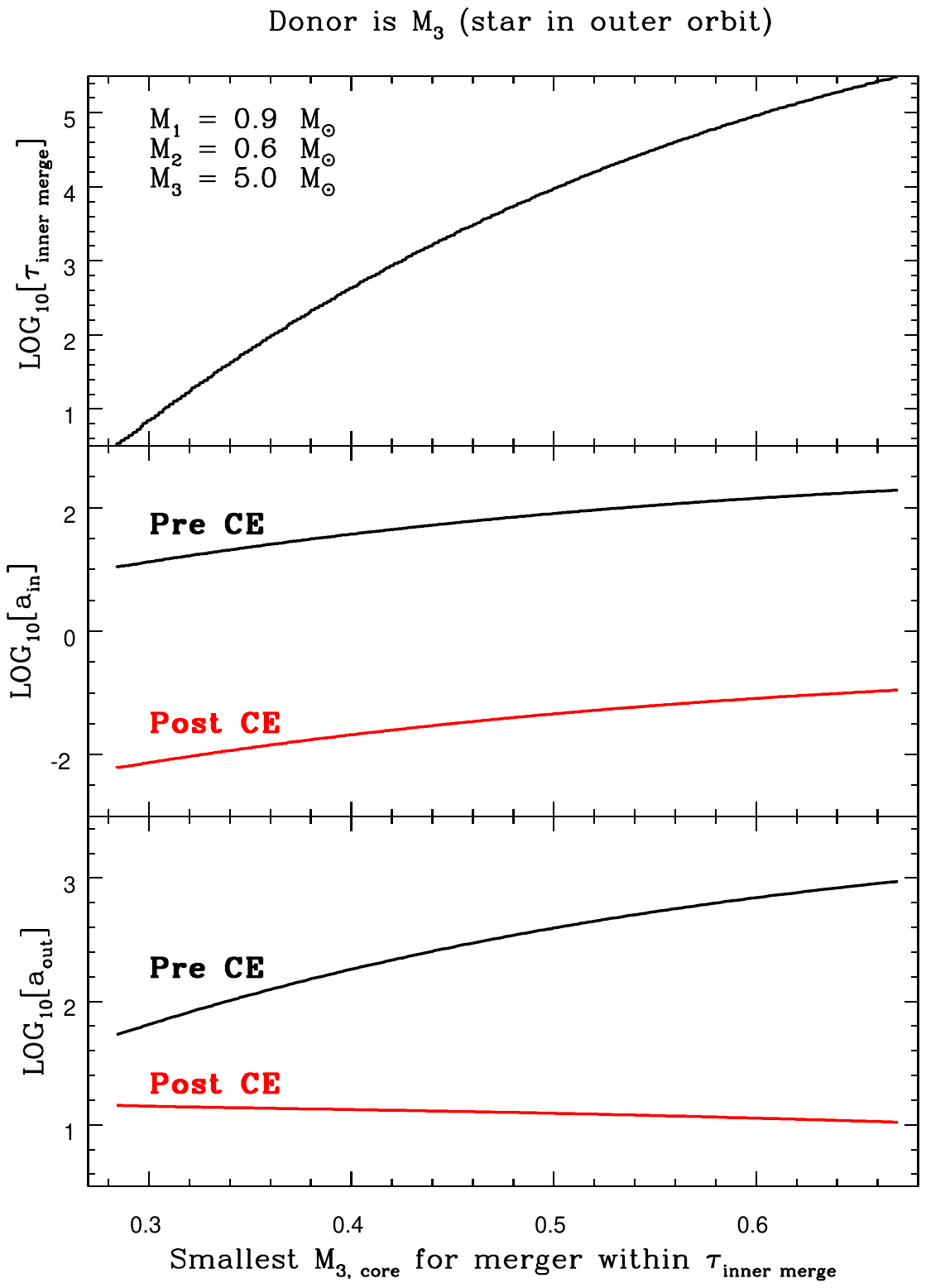}
\vspace{-.4 true in}
    \caption{
    Triple-CE results versus core mass of Star~3, the outer-orbit star that serves as donor. Each point on each curve represents a system in which the inner binary would not merge within a Hubble time, but which will
    merge sometime after a CE phase triggered by the RL filling of Star~3. {\bf Top panel:}
    the logarithm of the time (in units of the Hubble time) 
    required for the inner binary to merge without the intervention of a CE is plotted versus the minimum core mass needed to satisfy all of the conditions described in the text. {\bf Middle panel: } The pre-CE and post-CE orbital separations of the inner binary are plotted versus the core mass. {\bf Bottom panel:} The pre-CE and post-CE orbital separations of the outer binary are plotted versus the core mass.}
    \label{fig:cm3}
\end{figure}

One of the most important effects to occur when a star in the outer orbit fills its RL is that the inner binary can be driven to merger. We focus on this effect in Figure~\ref{fig:cm3}, which shows the final separations of the inner and outer binaries for a set of systems in which the outer star, with mass $M_3= 5\, M_\odot$, fills its RL; the inner binary consists of two WDs, with $M_1=0.9\, M_\odot$ and $M_2=0.6\, M_\odot$. We uniformly select the inner-orbit separation from a logarithmic distribution.
We then select a core mass for Star~3, with the masses ranging from $0.25$ to unity. Each value of the core mass corresponds to a value of Star~3's radius, which in turn translates into a value for the size of the outer orbit. We continue to consider only initial separations for which the three-body system is dynamically stable, and for which the inner binary will not merge in a Hubble time.  
Shown in the top panel of Figure~\ref{fig:cm3} are systems that will merge within a Hubble time as long as the core mass is larger than the mass given along the horizontal axis. The times, which are the times needed to merge if there were no CE, are expressed in units of the Hubble time, and the masses are expressed in units of a solar mass. 
None of these systems would have merged without the CE contributed by the third star.  

The middle and bottom panels show the pre-CE and post-CE orbital separations (in units of $R_\odot$) for the inner binary and outer binary, respectively. The values of the inner binaries' separations are consistent with mergers, while the outer binaries will not merge within a Hubble time, unless forces other than pure gravitation are brought to bear.

\section{Hierarchical Triple: An Inner Star Fills its RL}{\label{sec:5}}

We consider the case in which an inner star, Star~1, in a hierarchical triple fills its RL, creating a CE. If the third star is too distant from the inner binary, it may not exchange a significant amount of angular momentum with the envelope. We can then treat the inner binary as if it were isolated, employing Equation~\eqref{eq:afin_ratio}.  
The outer orbit is likely to widen due to the loss of mass from the inner binary.
We begin in Section \ref{sec:5.1} by identifying systems in which the third star is expected to be engulfed by the CE and to exchange angular momentum with it. 

\subsection{When Does the CE Engulf the Outer Star?}{\label{sec:5.1}}

Consider Star~1 in the inner binary. Its RL with respect to Star~2, its closest companion, is $R_{L,\mathrm{in}} = f(M_1/M_2)\times a_{\mathrm{in}}$.
If Star~1 fills this RL and if the conditions for dynamically unstable mass transfer apply, it will create a CE that engulfs both Star~2 and its own core, whose mass is $M_1^c$. If there is a third star it may also be engulfed by the CE. When, however, the value of $a_{\mathrm{out}}$ is too large, Star~3 will not exchange significant amounts of angular momentum with the envelope, and we can treat the inner binary simply by employing Equation~\eqref{eq:afin_ratio} with Star~1 and Star~2. In this case, the orbit of the outer star is affected primarily by the inner binary's loss of mass, and is most likely to expand.

The question we address below is: how small must $a_{\mathrm{out}}$ be, in comparison to the other sizes in the triple, in order for the star in the outer orbit to be engulfed by the CE?
We first consider the Roche geometry of the 
inner orbit. Star~1 fills its RL with respect to Star~2:
\begin{equation}
    R_1 = f(q_{1,2}) \times a_{\mathrm{in}} = R_{L,1}^{\mathrm{in}}.
\end{equation}
Here, $R_1$ is the radius of Star~1 at the point of RL filling. $R_{L,1}^{\mathrm{in}}$
is the RL radius of Star~1 with respect to Star~2, its companion in the inner binary. 

The geometry of mass exiting the binary may be complex. We know that the $L_2$ point defines a region through which mass is easily funneled out of the inner binary. Generally, the effective radius of the region defined by $L_2$ is roughly twice as large as the RL radius of the donor. We define the effective radius of the inner binary, from the perspective of mass loss, to be 
\begin{equation}
    {\cal {R}}_{\text{eff}} = L \times R_{L,1}^{\mathrm{in}},
\end{equation}
where $L$ is a parameter of order unity that, for the sake of estimation, we will later set to the value ``2''. We aim to determine the circumstances in which it is appropriate to consider the CE encompassing an {\sl outer binary} consisting of Star~3 and a companion of mass $M_2+M_1^{c}$. The inner companion, $M_2+M_1^{c}$, is taken to be located at the center of mass of the two stars, $M_2$ and $M_1^c$, and to have an effective radius ${\cal R}_{\text {eff}}$.   

In order for Star~3 to be engulfed by the CE generated from within the inner binary, ${\cal {R}}_{\text{eff}}$ must be a significant fraction of the RL radius of the inner binary with respect to the outer star. This latter quantity is 
\begin{equation}
R_{L,\mathrm{bin}}^{\mathrm{out}} = f\!\left(\frac{M_{\mathrm{in}}(0)}{M_3(0)}\right) \times a_{\mathrm{out}}. 
\end{equation}
Stability requires that $a_{\mathrm{out}}$ be larger than $G\times a_{\mathrm{in}}$.
Thus, 
\begin{equation}
\frac{R_{\mathrm{L,bin}}^{\mathrm {out}}}{f(M_{\mathrm{in}}/M_3)} > G. 
\end{equation}
We define a parameter, ${\cal P}$, to be the fraction 
\begin{equation}
    {\cal P} = \frac{{\cal {R}}_{\text{eff}}}{R_{\text{L,bin}}^{\text{out}}} =
    \frac{L\times R_{L,1}}{a_{\mathrm{out}} \times f\!\left(\frac{M_{\mathrm{in}}(0)}{M_3(0)}\right)}. 
\end{equation}
If the value of ${\cal P}$ is close to unity, we expect that the star in the outer orbit will be engulfed by the CE. If, on the other hand, the value of ${\cal P}$ is small, then it is more likely that the outer orbit will be less affected by the CE generated within the inner binary.  

We now ask what the values of ${\cal P}$ can be. For larger values of $a_{\mathrm{out}}$, ${\cal P}$ becomes smaller, indicating that the CE is less likely to engulf Star~3. Physical considerations specific to each triple can, in principle, be used to compute ${\cal P}_{\min}$, the smallest value likely to be associated with a three-body CE. 
In this paper, we simply choose ${\cal P}_{\min}$ to be $0.3$. Dynamical and/or hydrodynamical simulations can be conducted to determine if the minimum value should be larger, or if it can be smaller.  

There is also a maximum value of ${\cal P}$, ${\cal P}_{\max}$, which is determined by the minimum value of $a_{\mathrm{out}}$. If ${\cal P}_{\max}$ is smaller than the likely value of ${\cal P}_{\min}$, then a three-body CE will not occur. 
The stability criterion allows us to derive a lower limit for $a_{\mathrm{out}}$, hence an upper limit for ${\cal P}$ is 
\begin{equation}
    {\cal P}_{\max} = \left(\frac{L}{G}\right)\, \left(\frac{f(M_1/M_2)}{f(M_{\mathrm{in}}(0)/M_3)}\right).
\end{equation}

\begin{figure}
\includegraphics[width=\columnwidth]{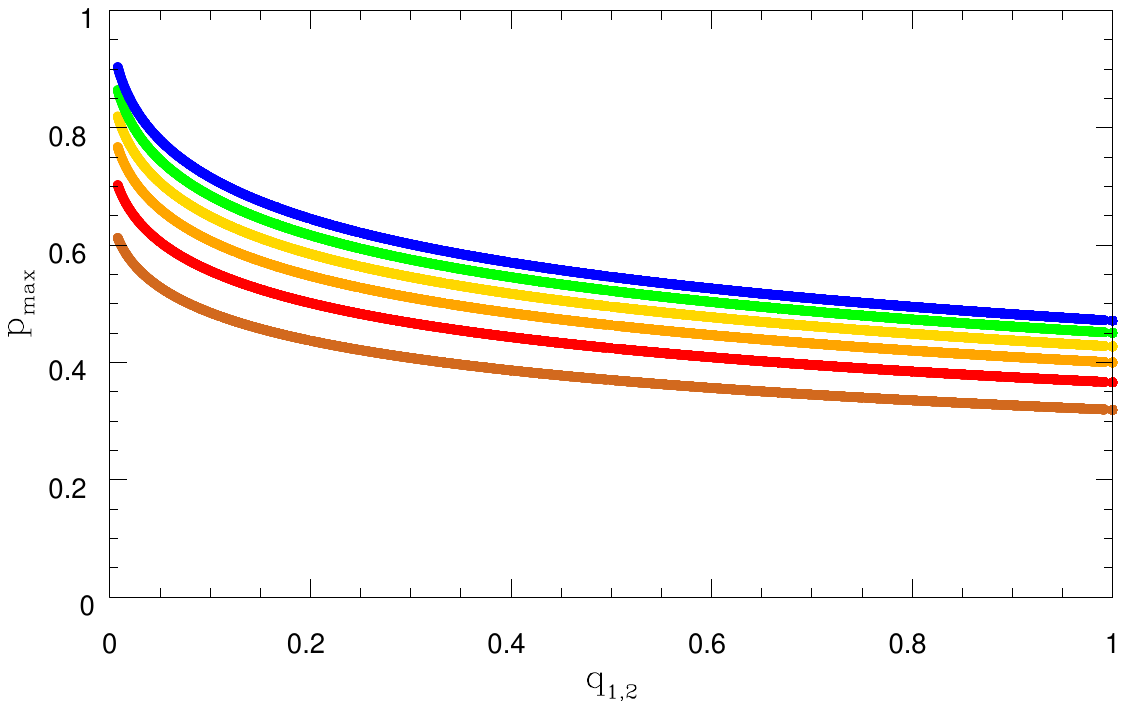}
\vspace{-2.5 true in}
    \caption{Values of ${\cal {P}}_{\max}$ are plotted versus $q_{1,2}$ for each of six values of $q_{3,{\mathrm bin}}$.
    The values of $q_{3,{\mathrm bin}}$ start at $1/3$ for the bottom curve and increase by $1/3$ for each curve above, reaching the value of $2$ for the topmost (dark blue) curve.}
    \label{fig:p_max}
\end{figure}

In Figure~\ref{fig:p_max} we show values of ${\cal {P}}_{\max}$ computed for a variety of systems. The calculations take Star~2 to be the RL-filling star; values of $M_2$ extend to $50\, M_\odot$. $M_1$ takes values up to $M_2$. 
Values of $q_{1,2}$ are plotted along the horizontal axis. Each curve corresponds to a single value of $q_{3,{\mathrm bin}}=M_3/(M_1+M_2)$. The bottom (chocolate-coloured) curve has $q_{3,{\mathrm bin}} =1/3$, and the value increases by $1/3$ for each curve above, reaching $2$ for the upper (blue) curve. The trend is clear: larger values of the outer mass yield larger values of ${\cal {P}}_{\max}$, as do small values of $q_{1,2}$. Most important is that the value of ${\cal {P}}_{\max}$ is larger than $0.4$ over a wide swath of the parameter space, telling us that, in many triple systems, the third star will participate in the CE. 

\subsection{Orbital Changes}

We consider the case in which the outer orbit is not too big, and show how the full three-body CE SCATTER formalism can be applied.    
When Star~1 fills its RL, it is convenient to write its mass as $M_1(0)=M_1^c+M_1^{\mathrm{env}}$, and $M_1(f) = M_1^c$. The masses $M_2$ and $M_3$ are constant.

\subsubsection{Changes in the Outer Orbit}

The outer binary consists of Star~3 in orbit with the center of mass of the inner binary. $M_{\mathrm{binary}}(0)=M_1(0)+M_2$; $M_{\mathrm{binary}}(f) = M_1(f) + M_2=M_1^c+M_2$.
The two ``stars'' spiralling toward each other within the CE are Star~3 and the ``core'', which consists of the inner binary: $M_{\mathrm{binary}}(f) =M_1^c+M_2$.
Because the inner binary is treated as a single mass, $M^{\mathrm{interact}}_{\mathrm{out}}= M^{\mathrm{env}}_1$, and we get
\begin{equation}
\begin{split}
\frac{a_{\mathrm{out}}(f)}{a_{\mathrm{out}}(0)}
=
\left(\frac{(M_1^c+M_2)+M_3}{[(M_1^c+M_1^{\mathrm{env}})+M_2] + M_3}\right)\\ 
\left(\frac{[(M_1^c+M_1^{\mathrm{env}})+M_2]}{M_1^{c}+M_2}\right)^2 \times Y,
\end{split}
\label{eq:18}
\end{equation}
where
\begin{equation}
    Y = \exp\left[- \frac{2\, M_1^{\mathrm{env}}}{(M_1^c+M_2)+M_3} \, {\cal{F}}\!\left(\frac{M_3}{M_1^c+M_2}\right)\right].
\end{equation}

\subsubsection{Changes in the Inner Orbit}

The inner orbit consists of the RL-filling star, Star~1, and Star~2. As above, $M_2$ is constant; $M_1(0)=M_1^c+M_1^{\mathrm{env}}$; while $M_1(f)=M_1^c$. While the envelope of Star~1 encompasses the entire three-star system, the portion of the envelope that interacts with the inner binary is
\begin{equation}
    M^{\mathrm{interact}}_{\mathrm{in}} = {\cal Q}\!\left(\frac{M_1^c+M_2}{M_3}\right)\times M_1^{\mathrm{env}},
\end{equation}
and we get
\begin{equation}
\frac{a_{\mathrm{in}}(f)}{a_{\mathrm{in}}(0)}
=
\left(\frac{M_1^c+M_2}{(M_1^c+M_1^{\mathrm{env}})+M_2}\right) 
\left(\frac{(M_1^c+M_1^{\mathrm{env}})}{M_1^{c}}\right)^2 \times Y,
\end{equation}
where
\begin{equation}
    Y = \exp\left[- \frac{2\, M^{\mathrm{interact}}_{\mathrm{in}}}{(M_1^c+M_2)} \, {\cal{F}}(M_1^c/M_2)\right].
\end{equation}

\subsection{Example}

\begin{figure}
\includegraphics[width=\columnwidth]{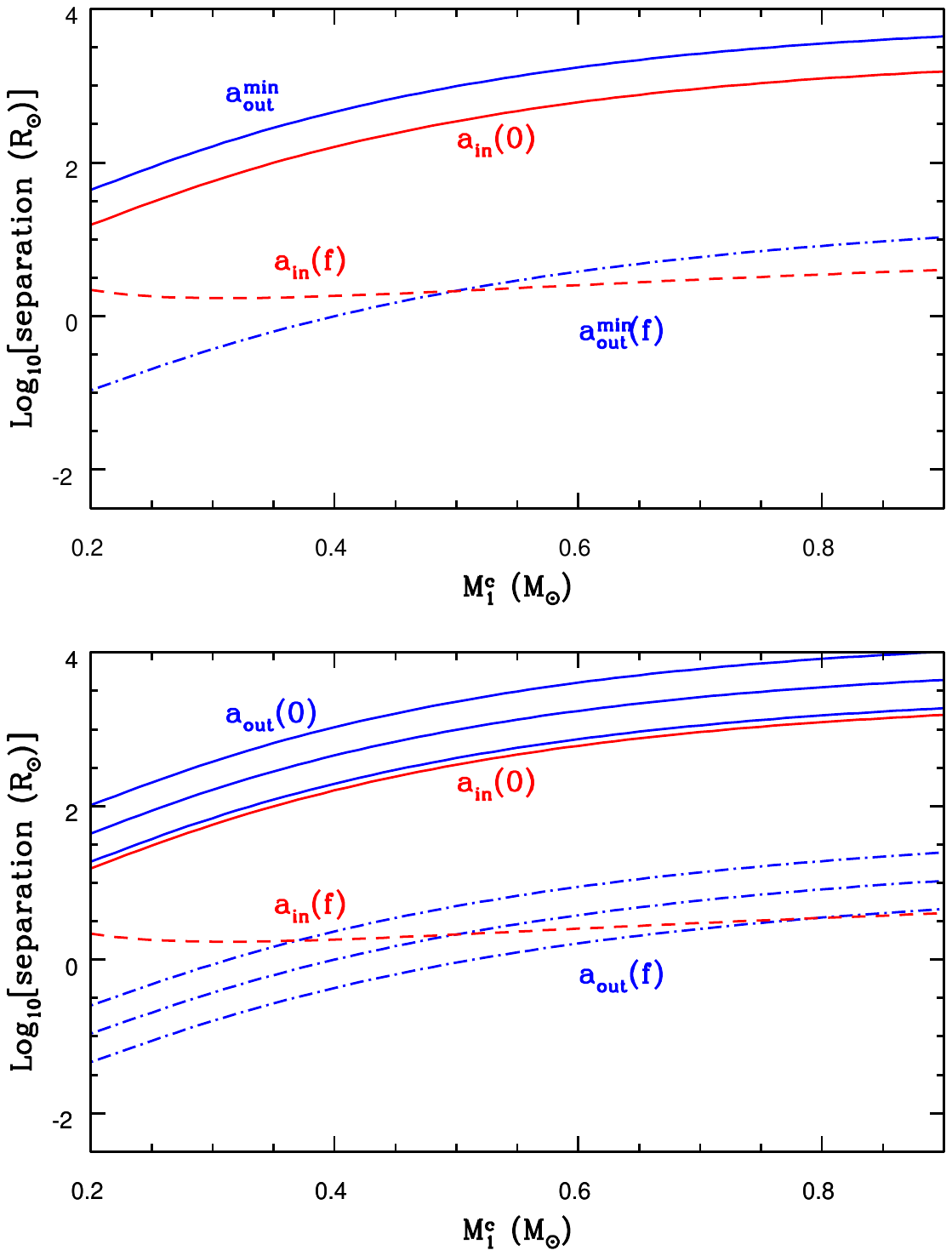}
\vspace{-.45 true in}
    \caption{The initial (pre-CE) orbit of Star~3 has radius $a_{\mathrm{out}}(0)$. Values of $a_{\mathrm{out}}(f)$ range from tens of solar radii, for the smallest core masses, to just under $10\, R_\odot$ for the largest possible core mass. The star occupying the post-CE orbit is a WD, with radius too small to merge with other stars, unless the other stars are large. In this case, the other stars are Star~1 and Star~2, whose orbital radius (i.e., the radius of the inner orbit) changes by a factor of $0.0003$, or even less. Thus, the inner binary will merge.}
    \label{fig:inner_fill}
\end{figure}

We consider the same trio of stars as in Section \ref{sec:4}, with masses $0.7\, M_\odot$, $0.83\, M_\odot$, and $5.0\, M_\odot$. We set $M_2= 0.7\, M_\odot$, as before, but exchange the labels of Star~1 and Star~3, so that $M_1$, the mass-giving star, is $5.0\, M_\odot$ and $M_3 = 0.83\, M_\odot$. The inner orbit consists of Stars~1 and 2, and the outer orbit consists of Star~3 orbiting the inner binary's center of mass. We vary the core mass of $M_1$ at the time of RL filling from $0.2\, M_\odot$ to $1.0\, M_\odot$. We compute ${\cal P}_{\max}$ to be $0.7$, indicating that the full three-body approach is required. 
The results are shown in the top panel of Figure~\ref{fig:inner_fill}, where we have taken $a_{\mathrm{out}}$ to have its minimum possible value, so that ${\cal P}= {\cal P}_{\max}$. 

For all values of Star~1's core mass, $M_1^c$, the inner binary shrinks to the size of a few solar radii. For cases in which Star~2 is extended, the core of Star~1 is likely to merge with Star~2, either during or shortly after the CE phase. 
The result will be a subgiant or giant, depending on the value of $M_1^c$, with an envelope comprised of a portion of the initial mass of Star~2. 
In cases in which both Star~1 and Star~2 are compact objects, they will merge after the CE, but within a Hubble time.  

This simple picture becomes more complicated, however, when the outer orbit is considered.
The outer binary shrinks so much that, for small values of $M_1^c$, it becomes significantly smaller than the inner binary. During the CE, Star~3's position becomes essentially the same as the position of the inner binary's center of mass. If both Star~2 and Star~3 are extended, all three stars will merge. The result will be a subgiant or giant with an envelope having a maximum mass $M_2+M_3$. The ultimate result will be a WD more massive than $M_1^c$. 

If just Star~2 is extended, then Star~3 is a WD. It will merge with the core of Star~1, within the cocoon consisting of the envelope of Star~2, which would itself be embedded within the remainder of the CE. 
If the sum of the masses of the two merging cores exceeds the Chandrasekhar mass, the result would be a SN Ia, occurring within an envelope with high central density. When $M_1^c$ is too small for the total merged mass to be above the Chandrasekhar mass, the two WDs would nevertheless merge, producing a highly energetic event taking place within the stellar and CEs. If just Star~3 is extended, the situation is exactly analogous.  

Note that, for the largest values of $M_1^c$, the outer orbit may have a post-CE size of a few $R_\odot$. If one or more of the stars is extended, the fates of the system would be the same as those described above. If all three are compact objects, they could merge within the CE. If, however, they do not, the result is a post-CE chaotic triple in which it is likely that all three compact objects will merge. The alternative is that two of the compact objects merge and the third star is ejected with high velocity. Given the masses considered in this example, any merger would be a merger of CO WDs, and for high values of $M_1^c$ all combinations potentially lead to SNe~Ia. A three-way merger would either produce a super-Chandrasekhar-mass SN~Ia or else an accretion-induced collapse to a NS.

In summary, the post-CE state of the system is almost certainly described by one of the following results. 
\begin{itemize}
\item An isolated fast-spinning giant or subgiant that eventually evolves to become a WD with mass greater than $M_1^c$. 
\item An energetic event in which two WDs merge within the envelope of a third star, with the CE still surrounding the system. The energetic event associated with the merger could be an SN~Ia or an AIC to a NS.
\item An energetic event or sequence of events in which three compact objects merge, leading to either a super-Chandrasekhar-mass SN~Ia or an AIC to a NS. The event may take place during or after the CE, but there will not be a surrounding envelope that was originally part of Star~2 or Star~3. 
\item The post-CE merger of two WDs, which could produce either a more massive WD or an SN~Ia, or an AIC, depending on the masses and compositions of the merging WDs. In addition a WD, either He–CO or CO, would be ejected from the system at high speed.
\end{itemize}

The bottom panel of Figure~\ref{fig:inner_fill} shows the results above, and also the results for cases in which the initial outer orbit has a different size from $a_{\mathrm{out}}^{\min}$. 
Near the top of the panel are three blue curves. The middle one of these is the same as shown for $a_{\mathrm{out}}^{\min}$, above. The top blue curve corresponds to an initial value of $a_{\mathrm{out}}$ that is $2.3$ times larger. For values of $M_1^c$ smaller than about $0.5\, M_\odot$ the results are the same as those discussed above. Larger values of $M_1^c$ are, however, associated with post-CE triples that have the potential to be stable.   

The bottom blue curve was derived by using the equations in this section. We note, however, that the initial outer orbit is too small to be consistent with dynamical stability. $a_{\mathrm{out}}(0)$ is $2.3$ times smaller than the stability limit. This system starts as a non-hierarchical triple. Note that all three curves for $a_{\mathrm{out}}(f)$ (including the top two, in which the triple is hierarchical) show that, for small values of $M_1^c$, the final separation of the third star from the inner binary's center of mass is small. This indicates the possibility of three-body mergers. For the lowest curve, where the three-body stability requirement is violated, the values remain small over a wider range of values of $M_1^c$. Furthermore, even for the highest values of $M_1^c$, the inner and outer separations track each other. Thus, even if this system would be able to survive the CE without mergers, it would evolve into a post-CE state that is dynamically unstable. We will consider a similar system in the next section and compare the results derived here with those derived by considering the triple as non-hierarchical.  

\section{The Triple is not Hierarchical}{\label{sec:6}}

\subsection{Overview}
When the triple is not hierarchical, we do not expect it to be dynamically stable at the start of the CE. All things being equal, the population of CE triple-star systems should be dominated by systems that are stable in their pre-CE state, simply because stable states are longer lived. There are, however, situations in which dynamical stability may not be satisfied. These include systems that have recently undergone episodes of stellar and/or binary evolution. In addition, dense stellar environments, such as within globular clusters or near the centers of galaxies, promote interactions between stars that can alter orbits or even produce capture events. Such interactions may yield triples in which the interstellar separations are comparable to each other. Finally, tidal disruptions of a passing star by a binary system may produce the type of system we consider in this section.  It is also important to note that the conditions for dynamical stability may not be the only determinants of whether a triple-star state is long lived. This is because non-dynamical physical effects can be significant. For example, tidal forces and/or heavy winds may be in play. Thus, the conditions for stability may be altered.  Although we expect that non-hierarchical triples form at most a small fraction of all triples experiencing a CE, we consider them for the sake of completeness.

\subsection{Modeling the Non-Hierarchical Case}

 In non-hierarchical initial states, the three separations $a_{12}(0)$, $a_{13}(0)$, and $a_{23}(0)$ are on an equal footing. This means that, if we want to compute the final state, we have three unknowns. Conservation of angular momentum provides only a single equation\footnote{The three-dimensionality of the angular-momentum vector does not help.}.  
In the hierarchical cases, we handled this potential roadblock by separately considering an inner and an outer binary. There is no obvious way to make a similar simplification for non-hierarchical triples.  
Instead, we view the non-hierarchical triple as a triangle whose vertices are the positions of the three stars and whose sides are the separations $a_{12}(0)$, $a_{13}(0)$, and $a_{23}(0)$.  
We apply SCATTER to each side separately. The final state can likewise be represented by a triangle whose vertices are the stars, but the sides of this triangle are given by $a_{12}(f)$, $a_{13}(f)$, and $a_{23}(f)$. Although the values of these separations may not exactly mirror the values that would be achieved in nature, they nevertheless provide good indicators of whether mergers occur.  

The SCATTER formalism relies on a fundamental physical principle: the conservation of angular momentum. To employ angular-momentum conservation, we need to estimate the fraction of the CE with which each star interacts. In \cite{DiStefano_2023} and in this paper, the function $\mathcal{Q}_\delta(q)$ provides such an estimate. Its functional form is based on the Eggleton function and therefore is related to the RL size, which depends on both gravity and rotation. We expect this definition to be fairly robust for binaries and for those hierarchical triples that can be modelled as if they were independent binaries.

For the non-hierarchical case, we apply a similar approach, based on the RL size of each ``binary'' (corresponding to the three sides of our triangle). We define the following expressions for the fraction of the CE with which each binary interacts:
\begin{gather}
  Q_{(1,2)} = \frac{\bigl[f\!\bigl(M_1/M_2\bigr)\bigr]^\delta + \bigl[f\!\bigl(M_2/M_1\bigr)\bigr]^\delta}{Q},\\
  Q_{(1,3)} = \frac{\bigl[f\!\bigl(M_1/M_3\bigr)\bigr]^\delta + \bigl[f\!\bigl(M_3/M_1\bigr)\bigr]^\delta}{Q},\\
  Q_{(2,3)} = \frac{\bigl[f\!\bigl(M_2/M_3\bigr)\bigr]^\delta + \bigl[f\!\bigl(M_3/M_2\bigr)\bigr]^\delta}{Q},
\end{gather}
where $Q$ is the sum of the three numerators, so that $Q_{(1,2)} + Q_{(1,3)} + Q_{(2,3)} = 1$.  

The quantity $Q_{(i,j)}$ represents the fraction of the envelope mass that exchanges angular momentum with the binary formed by stars $i$ and $j$. We apply SCATTER to each binary in the manner of \cite{DiStefano_2023}. We first compute the fractional change in angular momentum for binary $(i,j)$ by considering the contributions from stars $i$ and $j$ individually:
\begin{gather}
  \frac{dL_{(i,j),i}}{L_{(i,j)}} = \eta_{(i,j),i}\,
     \mathcal{Q}_\delta\!\left(\frac{M_i}{M_j}\right)
     \frac{dM_{(i,j),\mathrm{int}}}{M_i}\,
     \frac{M_j}{M_i + M_j},\\[4pt]
  \frac{dL_{(i,j),j}}{L_{(i,j)}} = \eta_{(i,j),j}\,
     \mathcal{Q}_\delta\!\left(\frac{M_j}{M_i}\right)
     \frac{dM_{(i,j),\mathrm{int}}}{M_j}\,
     \frac{M_i}{M_i + M_j}.
\end{gather}
The total fractional change in angular momentum is then
\begin{equation}
  \frac{dL_{(i,j)}}{L_{(i,j)}} =
    \frac{dL_{(i,j),i}}{L_{(i,j)}} +
    \frac{dL_{(i,j),j}}{L_{(i,j)}}.
\end{equation}
As in the binary case of \cite{DiStefano_2023}, the corresponding change in separation is
\begin{equation}
  \frac{da_{(i,j)}}{a_{(i,j)}} =
    2\,\frac{dL_{(i,j)}}{L_{(i,j)}} +
    \frac{d(M_i + M_j)}{M_i + M_j} -
    2\,\frac{dM_i}{M_i} -
    2\,\frac{dM_j}{M_j}.
\end{equation}

Now let us take our stellar components to be $1$, $2$, and $3$, with Star~1 donating an envelope of mass $M_1^{\mathrm{env}}$ and possessing a core mass $M_1^{c}$. To simplify the equations, we assume, following \cite{DiStefano_2023}, that $\eta_{(i,j),i} = \eta_{(i,j),j} \equiv \eta_{ij}$. Under this assumption, we obtain the following expressions for the change in separation of each ``binary'' in our non-hierarchical triple.
\begin{align}
\frac{a_{12}(f)}{a_{12}(0)} &=
  \left(\frac{M_1^c+M_2}{M_1^c+M_1^{\mathrm{env}} + M_2}\right)  \left(\frac{M_1^c+M_1^{\mathrm{env}}}{M_1^c}\right)^2  \notag \\[2pt]
&\quad \times
  \exp\!\Biggl[
    -2 \eta_{12}\,
    \frac{M_1^{\mathrm{env}}\,Q_{(1,2)}}{M_1^{c} + M_2}\,
    \mathcal{F}\!\bigl(M_1^{c}/M_2\bigr)
  \Biggr],
  \label{eq:a12}\\[6pt]
\frac{a_{13}(f)}{a_{13}(0)} &=
  \left(\frac{M_1^c+M_3}{M_1^c+M_1^{\mathrm{env}} + M_3}\right) \left(\frac{M_1^c+M_1^{\mathrm{env}}}{M_1^c}\right)^2 \notag \\[2pt]
&\quad \times
  \exp\!\Biggl[
    -2 \eta_{13}\,
    \frac{M_1^{\mathrm{env}}\,Q_{(1,3)}}{M_1^{c} + M_3}\,
    \mathcal{F}\!\bigl(M_1^{c}/M_3\bigr)
  \Biggr],
  \label{eq:a13}\\[6pt]
\frac{a_{23}(f)}{a_{23}(0)} &=
  \exp\!\Biggl[
    -2 \eta_{23}\,
    \frac{M_1^{\mathrm{env}}\,Q_{(2,3)}}{M_2 + M_3}\,
    \mathcal{F}\!\bigl(M_2/M_3\bigr)
  \Biggr].
  \label{eq:a23}
\end{align}

Here $\mathcal{F}(q)$ is defined as in Equation~\ref{eq:bigF}. In Appendix~\ref{appendix:nonhierarchical} we show that the set of equations would be the same if we first considered that each of the three stars exchanges angular momentum with a fraction of the envelope $Q_i$, and then considered the fraction of that mass exchanging angular momentum with one of its companion stars. 

\begin{figure}
\includegraphics[width=\columnwidth]{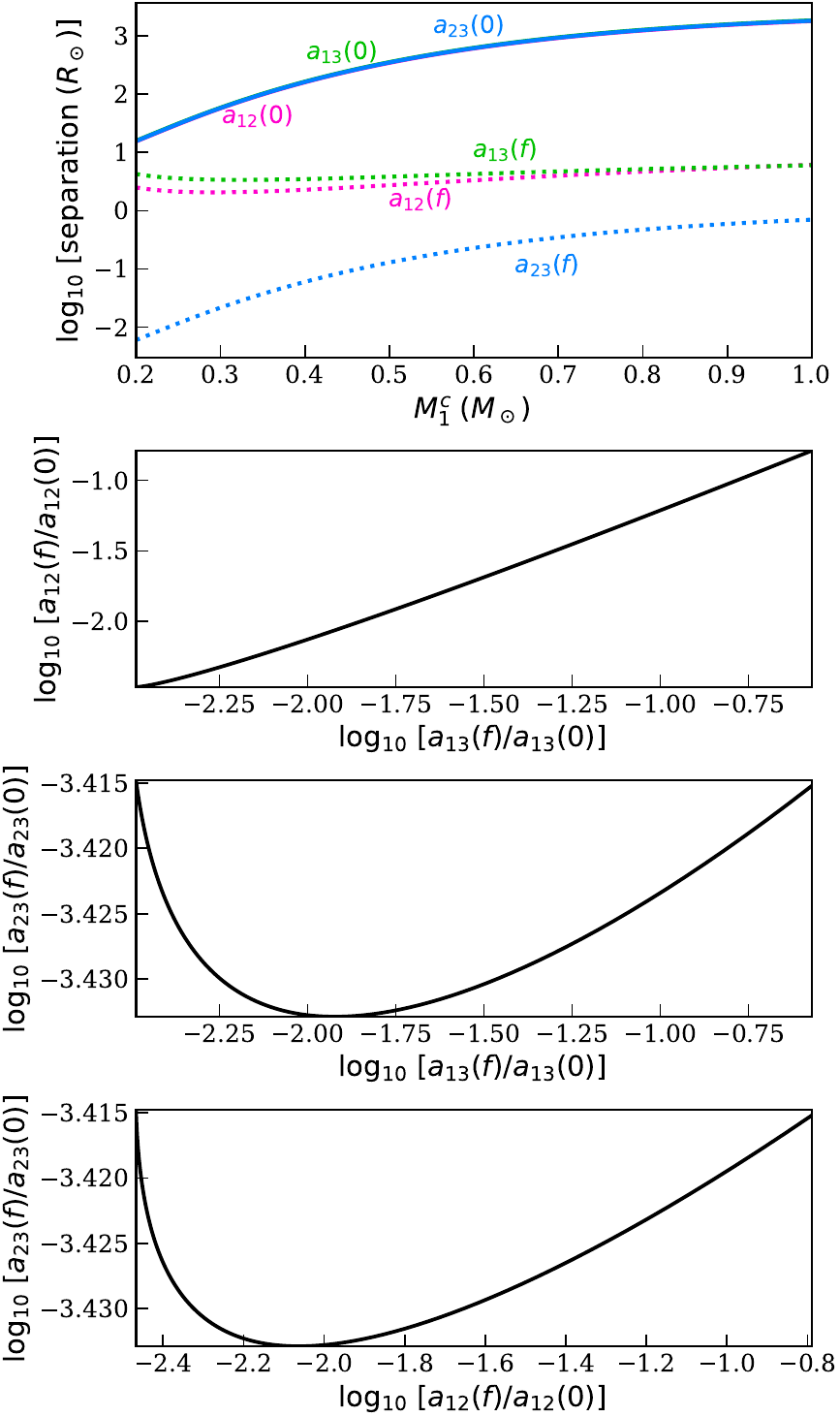}
\vspace{-.15 true in}
    \caption{The initial (post-CE) orbit of Star~3 has radius $a_3(0)$ ($a_3(f)$). Values of $a_3(f)$ range from tens of solar radii, for the smallest core masses, to just under $10\, R_\odot$ for the largest possible core mass. The star occupying the post-CE orbit is a WD, with radius too small to merge with other stars, unless the other stars are large. In this case, the other stars are Star~1 and Star~2, whose orbital radius (i.e., the radius of the inner orbit) shrinks by a factor of $0.0003$, or even less. Thus, the inner binary will merge.}
    \label{fig:nonhier}
\end{figure}

\subsection{Example}

We consider the same three stars used in Section 5, but start with a configuration in which the distances between each pair are the same, forming an equilateral triangle. 
Figure~\ref{fig:nonhier} illustrates the results. As might be expected, $a_{12}(f)/a_{12}(0)$ has a similar functional form with respect to $M_1^c$ to that of $a_{13}(f)/a_{13}(0)$. 
In this particular example, each separation, $a_{12}(f)$ and $a_{13}(f)$, is on the order of a few solar radii. There is likely to be a three-body merger if either Star~2 or Star~3 is an extended (non-compact) star. If not, the triple that results post-CE would not be hierarchical and could produce a merger and could also eject a high-velocity star. 
To consider the future of the system as a whole, we must take into account the distance between Star~2 and Star~3; $a_{23}$ decreases even more dramatically than the other distances, making it seem likely that all three stars will merge during the CE. 

These results suggest that Stars~2 and 3 will almost certainly merge, unless one or both suffer mass loss, or unless the value of $\eta_{23}$ is negative. $a_{12}$ and $a_{13}$ track each other, with Star~1 staying a few solar radii from both Stars~2 and 3. It is therefore possible that the post-CE state will be a close binary in which the core of Star~1 orbits the merger remnant of Stars~2 and 3. If so, the distance of closest approach is such that we expect a merger of the resulting binary within a Hubble time. If the result of the merger of Stars~2 and 3 is an extended star, then Star~1 is likely to merge with it even before the CE phase ends. Note that these results, particularly the high probability of mergers, are similar to the results derived in the hierarchical case, in which a star in the inner binary fills its RL, and the initial separation of the outer binary is marginally inconsistent with three-body dynamical stability. Thus, we can use the hierarchical case as a testbed to check against the results of the non-hierarchical case. 

Even if all of the stars merge within the CE, they may do so during two different events, each taking place within the CE. Thus, if an SN~Ia occurs during one event, the remaining star will be flung from the system with a potentially high velocity. If an SN~Ia does not occur before the second merger, the result could be either an AIC to a NS or else a super-Chandrasekhar-mass SN Ia. 

In making predictions for the non-hierarchical case, we must keep in mind that the model we have used may not perfectly reflect what happens in nature. We therefore benefit from being able to make comparisons with the case in which the model is that in which an inner star fills its RL in a binary with a smaller starting value of $a_{\mathrm{out}}$ than is consistent with three-body dynamical stability. As we have seen in \S~\ref{sec:5}, Figure~\ref{fig:inner_fill}, the results are similar.
The bottom three panels of Figure~\ref{fig:nonhier} show the logarithm to the base 10 of the ratios $a(f)/a(0)$ for each pair of stars, plotted against each other. 

\section{Stability of SCATTER'S Triple-Star Formalism}{\label{section7}}

\subsection{Stability}
No CE formalism can be relied on to make exact predictions.
They do aim, however, to correctly identify those systems which will either merge or experience future epochs of interaction. While the formalism may not give the same result as does nature for each system, the distribution of results should mirror what happens when real triple-star systems experience a CE. 
One way to test the formalism is to study its stability with respect to parameter changes. Two classes of parameters require study.

\begin{enumerate}
    \item {\bf System parameters:} When the formalism predicts a particular fate for a system, it should not be the case that small changes in the characteristics of the triple, e.g. masses or mass ratios, produce radically different results. While the end state should change as the characteristics of the triple-star system change, the alterations should be gradual. The quantity we calculate is $\frac{a(f)}{a(0)}$ for binaries within the triple. To illustrate the dependence of the formalism on system parameters we note that this ratio can be expressed in terms of mass ratios: $q_{12} = \frac{M_1(f)}{M_2(f)}$, $q_{13} = \frac{M_1(f)}{M_3(f)}$, $q_{23} = \frac{M_2(f)}{M_3(f)}$, and $q_{\mathrm{ec}} = \frac{M^{\mathrm{env}}}{M^c}$. Furthermore we can note that since $q_{13} = q_{12} \times q_{23}$, the SCATTER formalism for triples is only dependent on at most three unique mass ratios which describe the CE forming system. The plots within this section serve to illustrate the dependence of the CE on these parameters.
    \item {\bf Formalism parameters:} A number of assumptions are made in each CE formalism. The basic physical assumption of SCATTER is simple: angular momentum drawn from each component of the triple is responsible for the change in that component's orbital angular momentum. To implement this simple requirement requires defining several quantities. Perhaps the primary example is the design of the functions that allow us to compute the fraction of the envelope with which each component interacts. These functions have been denoted by the letter ${\cal Q}$ and, for non-hierarchical triples, $Q_{(i,j)}$. The choices we have made are physically reasonable. In addition, we note that the choice of ${\cal Q}$ for binaries is fully consistent for all known post-CE WD-containing systems. Nevertheless, the collection of data on a wider range of post-CE systems could lead to a preference for different functions. 

    The other important choice in SCATTER is the functional form of $\eta$, which is an indication of the efficiency of angular-momentum transfer. In principle, each component that transfers angular momentum to or from the CE does so with an efficiency that could very well be different from the analogous efficiency of other components of the triple. Simulations can be designed to study this issue. In the absence of more data on post-CE systems, or of sound physical reasons to introduce additional parameters, we have assumed that $\eta$ is the same for all components. Comparisons with post-CE systems reveal that this key quantity is a function of mass ratios. For the WD systems we have considered, the function is linear in log space, leading to the introduction of the parameters $A$ and $B$, which are fit from the data. We can therefore study the stability of the formalism itself by varying the values of $A$ and $B$. We note of course that post-CE data on other systems (e.g., BH–BH systems) may lead to a non-linear functional form, thereby introducing additional parameters. For now, however, the clear way to test the stability of the formalism is to vary the values of $A$ and $B$. For WD-containing post-CE systems, we generally found that, when considering different groups of post-CE binaries, the value of $B$ was correlated with the value of $A$, a circumstance we use in producing our graphs.   
\end{enumerate}

Figures~\ref{fig:sc2input} and \ref{fig:ainslices} describe hierarchical triples; those in Figures~\ref{fig:a12slices}--\ref{fig:a23slices} describe non-hierarchical triples.
 The contours in these plots each represent a specific value of $\log_{10}[a(f)/a(0)]$, which we will call the shrinkage factor. To determine whether there is a merger, we need to consider particular systems with known values of $a(0)$; we also need to know the radii of the stars. Figures illustrating the shrinkage factors by themselves are useful. Consider a shrinkage factor of $10^{-2}$. If the separation between the stars was initially in the range of a few $R_\odot$ to about a hundred $R_\odot$, we would expect a merger. For larger initial separations we might have mergers, depending on the radii of any extended stars in the triple. There is also the possibility of a post-CE merger and/or other post-CE interactions. Note that, within the SCATTER formalism, binary expansion also occurs in some cases.

A shrinkage factor below $10$ will significantly affect the future interaction of stars that started the CE as main-sequence stars or as subgiants. Even if the RL-filling star is a giant, a shrinkage factor of 10 could bring its core close enough to the inner components of the triple (which may have merged) to facilitate future interactions, particularly if one of the other stars is extended.  

\subsection{Star in the Outer Orbit of a Hierarchical Triple Fills its RL}

We first consider the scenario in which the outer star in a triple system acts as the envelope donor. In this case, the SCATTER equations are formulated as functions of two distinct mass ratios, one which describes the mass ratio of the components of the inner/outer binary, and the other which is tied to the interacting envelope mass. For the outer binary these two ratios are $q_{\mathrm{out}} = \frac{M_3^c}{M_{\mathrm{binary}}} = \left(q_{13}+q_{23}\right)^{-1}$, and $q_{\mathrm{ec}} = \frac{M_3^{\mathrm{env}}}{M_3^c}$. For the inner binary they are $q_{\mathrm{in}}=q_{12}$, and $q_{\mathrm{ebin}} = \frac{M_{{\mathrm{in}}}^{\mathrm{interact}}}{M_{\mathrm{binary}}}=\frac{q_{\mathrm{ec}}}{q_{13}+q_{23}} {\cal Q}\!\left(\frac{1}{q_{\mathrm{out}}}\right)$.

\begin{figure*}
\includegraphics[width=18.5cm]{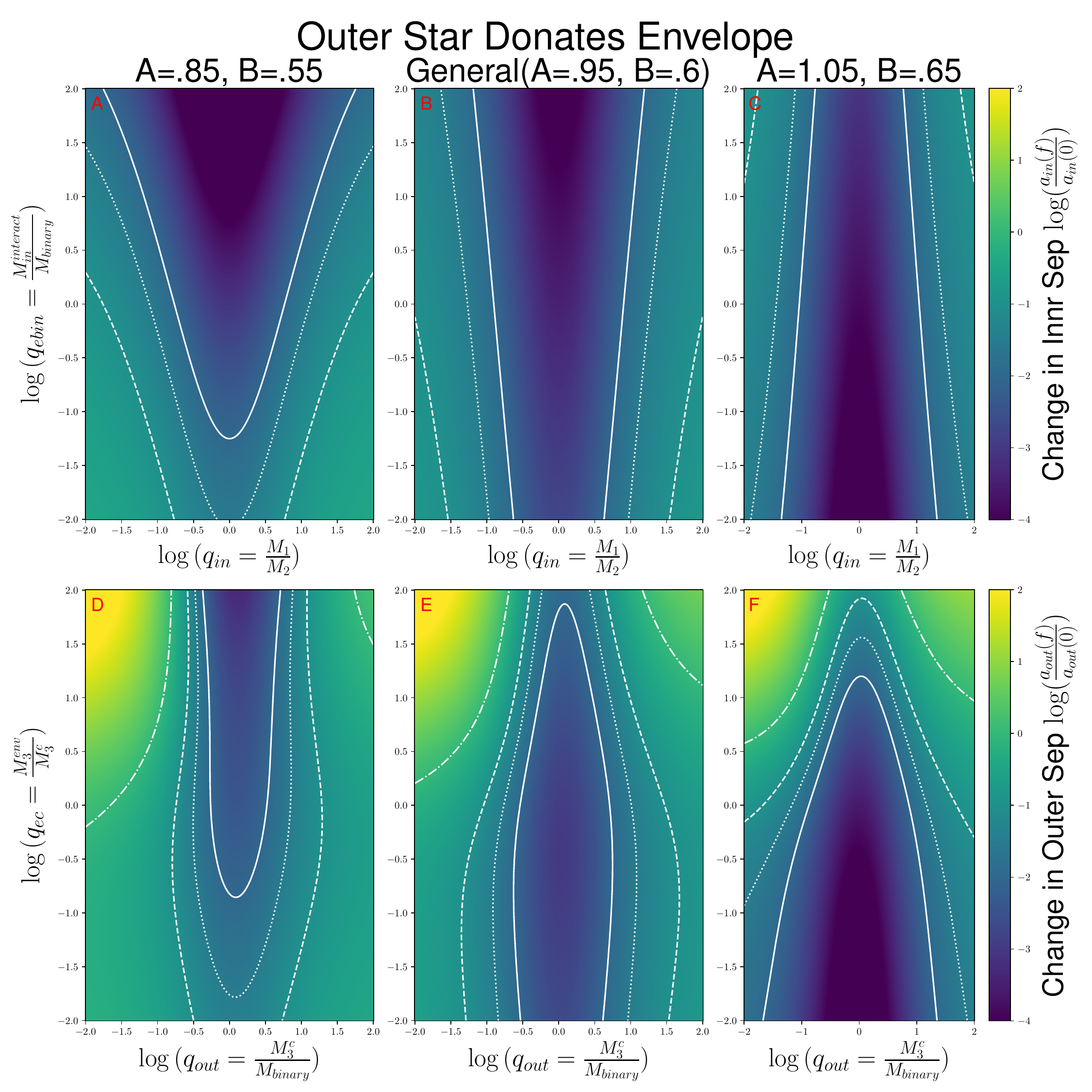}
\vspace{-.1 true in}
\caption{Colour plot illustrating the logarithmic change in outer separation for both inner (top panels) and outer (bottom panels) binaries when the outer star acts as the donor. We generate three sets of plots corresponding to different fitted values of $\eta$ as provided by \protect\cite{DiStefano_2023}, to cover a representative range of angular-momentum transfer efficiency. These values are then plugged into our SCATTER equations for the inner and outer binary, as described by Equations~\ref{eq:7} and \ref{eq:10}. The solid white contour represents shrinkage factors beyond $10^{-2}$, signifying conditions where mergers are nearly inevitable. The remaining three contour lines indicate shrinkage factors of $10^{-1.5}$, $10^{-1}$, and unity, respectively.}
\label{fig:sc2input}
\end{figure*}

The upper panels of Figure~\ref{fig:sc2input} show that, for all of the mass-ratio values we have considered, the inner binary always shrinks. Note that the ranges of mass ratios in Figure~\ref{fig:sc2input} extend far beyond those considered in the examples of previous sections, and even beyond those that would be achieved in a broad swath of triple systems of high scientific interest. For values of the ratio of the envelope mass to the binary mass above unity, the effect of increasing $A$ is to create more shrinkage, while the opposite is true for small values of the ratio (which is plotted along the vertical axis).

In all cases the amount of shrinkage is significant, and will lead to mergers and/or to subsequent interactions of the components of the inner binary across a broad sample of physical systems. Furthermore, changes in the amount of shrinkage occur gradually as the mass ratios shown along both the horizontal and vertical axes change. 

The three plots in the upper panel also demonstrate stability with respect to the formalism's input function, $\eta$. To vary $\eta$, we vary the values of $A$ and $B$. As $A$ and $B$ change, the regions in the $q_{\mathrm{ebin}}$–$q_{\mathrm{in}}$ plane with maximum shrinkage can be seen to shift. The shift is gradual and well regulated. 

The figures in the bottom panel correspond to the amount of outer-orbit shrinkage produced by the CE. The new feature in the bottom panels is that the points in colours ranging from light green to yellow illustrate that the outer binary can expand; the shrinkage factor can be negative. This is because the outer orbit can be as influenced by the loss of matter from the system as it is by the loss of angular momentum to the CE. This also explains the more modest amount of shrinkage, relative to the inner binary, generally seen in the lower panels. 

A primary trend observed in Figure~\ref{fig:sc2input} is that orbital shrinkage is maximized when the binary components possess equal masses. Mathematically, this corresponds to the fact that the function $\mathcal{F}(q)$ is maximized at $q=1$ (see Figure~2 of \citealt{DiStefano_2023}). Nevertheless, there are broad ranges of mass ratios, centred around $M_3^c/M_{\mathrm{bin}}$ and across a broad range of values of $q_{\mathrm{ec}}$, where mergers and/or subsequent episodes of interaction are predicted.

For the inner binary, separation consistently decreases due to the CE event, because the only change in angular momentum comes from the CE exerting torque on the binary. Conversely, the outer binary experiences an additional angular-momentum shift arising from mass loss of the third star, leading to scenarios in which orbital separations may increase post-CE. These results should be viewed as average trends. If, for example, there is matter ejected from the inner binary, it may not shrink as much or may even expand.

Furthermore, we identify two distinct regimes characterized by the dependence on the envelope–core mass ratio ($q_{\mathrm{ec}}$ for the outer binary and $q_{\mathrm{ebin}}$ for the inner binary). The first regime, dominant at lower $A$ values, is primarily governed by the mass of the envelope. Here, higher values of $q_{\mathrm{ec}}/q_{\mathrm{ebin}}$ correspond to more massive envelopes, leading to more torque on the components of the triple, thereby increasing the likelihood of mergers. This is illustrated by the solid contours in the panels A, B, and D.

At higher $A$ values, the second regime emerges, in which the parameter $\eta$ (inverse angular-momentum transfer efficiency) decreases rapidly with higher envelope–core mass ratios. As angular-momentum transfer efficiency improves in this regime, fewer mergers occur at elevated ratios, a behaviour evident in the shape of the white contours in the panels C, D, and F.

\subsection{Star in Inner Orbit Fills its RL}

Similar to the outer-binary case, the equations describing the inner binary can also be expressed as functions of mass ratios. Furthermore it is easy to see that the case of the outer binary is functionally identical to the case where the outer star is the donor as one can see by noting that Equation~\ref{eq:7} is the same as Equation~\ref{eq:18} but with $M_3^c \to M_3$, $M_1 \to M_1^c$, and $M_3^{\mathrm{env}} \to M_1^{\mathrm{env}}$. We therefore do not show the temperature plots for the outer binary. 

The inner-binary scenario is now dependent on all three unique mass ratios which represent the system, as it both loses mass and shares its envelope with the outer star. We examine the interplay among these three ratios in further detail in Figure~\ref{fig:ainslices}.

The general dependence of orbital separation on $q_{12}$ and $q_{\mathrm{ec}}$ along with $A$ and $B$ for an inner envelope donor closely parallels the behaviour observed when the outer star is the envelope donor, reflecting the consistent underlying theoretical framework. The primary distinction in the inner-binary scenario arises from the additional effect of $q_{23}$. At lower $q_{23}$ values, the inner binary receives a progressively smaller fraction of envelope mass, thereby reducing angular-momentum loss during the CE phase as seen in the leftmost panels of Figure~\ref{fig:ainslices}. Consequently, in this range of $A$ and $B$ values, significant shrinkage of the inner-binary orbit predominantly occurs when the combined mass of the donor star's core and its companion star is similar to or larger than the mass of the third star. This trend changes for higher values of $A$ and $B$ as seen in the rightmost plot of these figures where shrinkages are minimized at intermediate values of $q_{23}$. This is because, as in the previous case, at higher $A$ values we transition into an $\eta$-dominated regime where the high value of $\eta$ corresponds to greater shrinkages of the inner orbit and thus mergers over a much wider portion of the parameter space. The exception to this trend is in the case of very large envelope–core mass ratios where high values of $q_{\mathrm{ec}}$ lead to significantly lower values of $\eta$ and thus less shrinkage of the inner orbit.

Additionally we recall that due to the critical mass-ratio criterion for CE formation, scenarios characterized by small values of $q_{23}$ will only be realized for higher values of $q_{12}$ and $q_{\mathrm{ec}}$. Thus the realized portion of the parameter space is generally smaller than that over which we choose to plot our figures. These plots highlight the complexity and diversity of outcomes possible for inner-binary orbital evolution.

\subsection{Star in Non-Hierarchical Triple Fills its RL}

We now turn to the scenario in which the initial triple configuration is non-hierarchical. This implies that the three interstellar separations --- $a_{12}$, $a_{13}$, and $a_{23}$ --- are of comparable size, and no clear inner or outer binary exists. To address this, we treat all three pairs symmetrically, applying the SCATTER formalism independently to each ``binary'' as described in Equations~\ref{eq:a12}–\ref{eq:a23}.

Figures~\ref{fig:a12slices}--\ref{fig:a23slices} present the results for this configuration. Each panel shows the logarithmic change in orbital separations for one of the three component binaries --- (1,2), (1,3), and (2,3) --- as a function of the mass ratios $q_{12}$, $q_{23}$, and $q_{\mathrm{ec}}$ over a variety of different fits for $\eta$.

The form of our equations means that, for most systems, both binary (1,2) and binary (1,3) are functionally rather similar to the hierarchical case in which a star in the inner binary donates its envelope. In Figures~\ref{fig:a12slices} and \ref{fig:a13slices} we see similar trends over our three main parameters as in Figure~\ref{fig:ainslices}, showing the natural nature of the extension of the formalism to non-hierarchical systems. Furthermore we note that the equations for binary (1,2) and (1,3) are functionally identical but with $M_2$ and $M_3$ swapped, so that the two plots closely mirror each other. While these hierarchical and non-hierarchical cases are similar, the non-hierarchical triple case generally displays less shrinkage in the orbital separation between the donor star and its companions. Such a trend makes physical sense as the envelope mass is shared between binaries (1,2) and (1,3).

Binary (2,3) on the other hand represents a case more similar to that of the inner binary when a star in the outer binary donates its envelope. This is due to the fact that neither star loses mass; the CE thus leads only to shrinkage.

These results demonstrate that non-hierarchical triple systems undergoing a CE phase are likely to produce at least one merger, 
with some configurations leading to situations in which two or even all three of the separations decrease significantly during the CE.

\section{Conclusions}{\label{sec:8}}

\subsection{Formalism for Three-Body Systems}

We have developed a formalism to compute the results of CE episodes in triple-star systems. The method is stable in the sense that its predictions change gradually as the physical parameters of the initial three-body system gradually change. It is also stable, behaving in expected ways, to modifications of the input functions.  

When considering events in which spiral-in within a CE plays a role, triple systems can differ from binaries in two ways. First, whichever star generates it, the CE generally brings at least one pair of stars closer together. As illustrated by some of the specific examples we considered, some triple CEs can drive binaries to merger even if they would not otherwise have merged. Whether such binaries merge during the CE or after, the merger time is earlier than it would otherwise have been. The system is more likely to contribute to early-time events.

Second, the remaining star is often closer to the merged star than it otherwise would have been. The third star may be brought close enough to: {\bf (a)}~allow a future episode of stable mass transfer; {\bf (b)}~lead to a second CE episode; or, {\bf (c)}~merge within a Hubble time. Further interactions involving the third star can occur within or slightly after the CE, thereby contributing to early-time events. Further interactions may instead occur at late times, contributing to late-time events, thereby enhancing the rates over the full range of time scales. 
Even in cases where an orbit containing an unevolved star widens, the expansion may still yield a system in which further episodes of mass transfer are possible.  

\subsection{Predictions of Events and Post-CE States}

Common-envelope calculations, particularly in the realm of population synthesis, are used to compute a variety of event rates. The event types range from SNe~Ia to gravitational-wave mergers of BHs and/or NSs. Correctly predicting event rates is therefore a crucial goal of CE formalisms.  
Fully incorporating triples into the calculations is crucial to making correct predictions. In fact, triples are capable of producing more events per system than binaries. It is essential to include them because a large fraction of the events and end states we aim to explore emerge from primordial triples, which are common among massive stars.

If triple-inspired mergers are common, this relieves the pressure on binary models to produce all of the compact-object mergers observed.
In fact, higher-order multiplicity may be needed to produce the most massive mergers or mergers with other unusual features. These latter include unusual spin values or orientations, or the chemical output of mergers involving NSs. Since triples and other higher-order multiples open more channels to merger, they may even be the dominant formation channel for some systems. 

\subsection{Immediate Applications}

\subsubsection{Type~Ia Supernovae and Other WD Systems}

Because the fits to the $\eta$ function are based on post-CE systems containing WDs, SCATTER is ready for simulations designed to compute the rates and delay-times of SNe~Ia. With the triple formalism in hand, we can compare the rates and characteristics of events as derived in triple-star systems and in double-star systems.  

In the course of any simulation designed to compute the rates of SNe~Ia associated with various evolutionary channels, we can also compute the rates of formation of other WD-containing systems, including cataclysmic variables (CVs).

\subsubsection{Gravitational-Wave Mergers}
In binary scenarios, two BHs must form and then come close enough to each other to merge in a Hubble time. Triple-star systems provide additional opportunities. For example, a pair of BHs may not be close enough to each other to merge within a Hubble time. But if, after the BH binary has formed, there is still a star in an outer orbit, a CE induced by the evolution of that star may either cause the BHs to merge within the CE or else to be brought close enough during the CE that they will merge in a Hubble time. Furthermore, the remnant of the outer star could merge with the merged remnant of the inner star, as considered in \cite{stegmann_2022}.

\subsubsection{Higher-Order Multiples}

In this paper, we used the foundation provided by the binary-star SCATTER formalism to formulate methods for triples experiencing a CE. The extension to quadruples and other higher-order multiples can now be considered. As for both binaries and triples, we do not know exact values of the fraction of the mass interacting with each star, or the exact values of the angular-momentum-transfer efficiency. Nevertheless, straightforward approaches can be developed with the tools provided here to explore the possible numbers of mergers and other events, and to explore trends associated with different types of system.  

\subsection{Further Development of the Formalism}

The SCATTER two- and three-body formalisms are based on the fundamental principle of angular momentum conservation. Because each component of the system is allowed to interact with the CE, the formalism is flexible enough to produce a full range of effects, including orbital expansion. Furthermore, the functions that determine the fraction of the envelope interacting with each triple-star component, and the efficiency of angular-momentum transfer, are adjustable in a way that will accommodate input from future data and also comparisons with simulations. In Section \ref{section7} we have demonstrated that the results are stable with respect to changes in the efficiencies.

An important feature of the formalism is that the efficiency of angular-momentum transfer is determined by a function of the mass ratios in the triple system. This means that, within a single population-synthesis simulation, the efficiency factor changes from system to system.  

If future development of SCATTER shows that separate functions apply to different stars, the changes can be incorporated directly into individual population-synthesis simulations. This is a great advantage for allowing results that mimic nature to be derived within a single simulation. 

\begin{itemize}
    \item We have considered cases in which the CE mass is not accreted by any star in the system. While this is likely to be the case in many triples, there are situations in which a potential donor can accrete matter on a short timescale. The derivation of the SCATTER equations can then be derived in a straightforward way by using the basic angular-momentum conservation requirement with the inclusion of mass gain. 
    \item We have considered cases in which only the envelope-losing star ejects mass. If one of the other stars also loses mass, this can also be dealt with by using the basic angular-momentum conservation requirement.
    \item The functional form of $\eta$ was derived using data from observed post-CE systems. These systems all contained at least one WD. Some contained two WDs, others contained one WD and either a main-sequence or NS. In each case we found that $\log_{10}[\eta]$ is a linear function of $\log_{10}\!\left[M^{\mathrm{interact}}/M_{\mathrm{tot}}(f)\right]$. 
    We fit the slope, $A$, and the y-intercept, $B$, for each line and employed an average for our fiducial value in this paper. Data based on a larger number of post-CE systems and for different kinds of post-CE systems would refine the values of $A$ and $B$.
    \item In addition, data on post-CE states that do not contain WDs may produce more complex formulae for $\eta$, for example quadratic forms that would introduce another fit parameter.
    \item The formalism requires introducing a functional form for ${\cal Q}_i(q)$, the fraction of the envelope mass that exchanges angular momentum with component $i$. As pointed out in Section \ref{sec:etainner}, while we have made reasonable choices, they are not the only possible choices.
    \item We have considered circular orbits. While the population of triples includes systems in which an inner or outer binary may be eccentric, we do not explicitly include eccentricity in the CE calculations described in this paper.   Note, however, that  eccentricity could have played a role during the pre-CE evolution. It can also play a role once the CE has been dispersed.   It will be useful to extend the SCATTER formalism to be to take into account the role of eccentricity during the CE itself; it would circularize the system with an initial eccentricity assumed.  This will allow explicit consideration of a number of interesting cases.  For example, if the outer orbit is eccentric, the star in the outer orbit may plunge into a CE that encompasses the inner two stars.
\end{itemize}

The SCATTER formalism is constructed on the simple but powerful base of conservation of angular momentum. By allowing each star within a multiple-star system to interact with a portion of the envelope, the formalism provides a large parameter space that should allow post-CE results to reflect reality, at least in a statistical sense. Input parameterized functions, $\eta_i$ and ${\cal Q_i}$ can be modeled to reflect new data from observations and/or from numerical simulations. Parameter values that define system properties each yield values of the angular-momentum-transfer efficiency, allowing for consistency within population-synthesis codes. We have shown that the formalism has direct extensions to triples. Future work will derive results for a wide range of physical events. 

\section*{Acknowledgements}
The authors thank Yan Gao for discussions. RD and CK also thank Matthias Kruckow and Patrick Neunteufel for discussions.  
R.D. acknowledges support from the National Science Foundation, through NSF AST-2009520. 
C.K. acknowledges funding from the UK Science and Technology Facility Council through grant ST/Y001443/1.
We thank the referee, Jakob Stegmann, for helpful comments.

\section*{Data Availability}
Model data used in this paper can be shared on request. 

\appendix
\section{Alternative Derivation for Non-Hierarchical CE}{\label{appendix:nonhierarchical}}
In this approach instead of looking at the fraction of the CE which is associated with each of the three binaries, we instead determine the fraction of the CE attributed to each star in the triple. This then gives us the following three equations:
\begin{gather}
    Q_1 =  \frac{[f(\frac{M_1}{M_2})]^\delta + [f(\frac{M_1}{M_3})]^\delta}{Q},\\
    Q_2 = \frac{[f(\frac{M_2}{M_1})]^\delta + [f(\frac{M_2}{M_3})]^\delta}{Q},\\
    Q_3= \frac{[f(\frac{M_3}{M_1})]^\delta + [f(\frac{M_3}{M_2})]^\delta}{Q}.
\end{gather}
Where again value of  $Q$ is the same as in the primary derivation, and similarly we have that $Q_1+Q_2+Q_3=1$.
The value of $Q_i$ thus represents the fraction of the envelope mass that exchanges angular momentum with Star~$i$. In the non-hierarchical case we presently consider, we are also interested in the fraction of the envelope mass that interacts with each of the three binaries, $a_{12}, a_{13},$ and $a_{23}.$

Since we are considering three different binaries, we also devise a way to compute the fraction of the envelope mass that exchanges angular momentum with each binary.
To accomplish this in a consistent way, for each star i, we divide the fraction of the envelope with which it interacts into two portions: the portion that interacts with the binary $(i,j)$ and the portion that interacts with the binary $(i,k)$. 
For the sake of clarity we represent these fractions, respectively, as $\alpha_{i\to(i,j)}$ and $\alpha_{i \to {(i,k)}}$. We require $\alpha_{i\to(i,j)}+\alpha_{i \to {(i,k)}}=1$.  We then adopt a reasonable  functional form for $\alpha_{i \to (i,j)}$ where we consider the effective RL sizes of the two binaries of which Star~$i$ is a component:
\begin{equation}
    \alpha_{i \to (i,j)} = \frac{f\left(\frac{M_i}{M_j}\right)^\delta}{f\left(\frac{M_i}{M_j}\right)^\delta+f\left(\frac{M_i}{M_k}\right)^\delta}.
\end{equation}
From this we know that for a given star $i$ in binary $(i,j)$ the fractional change in angular momentum is given as
\begin{gather}
    \frac{dL_{(i,j),i}}{L_{(i,j)}} = \eta_{(i,j),i} \alpha_{i \to (i,j)} Q_i \frac{dM_e}{M_i} \frac{M_j}{M_i+M_j},\\
    \frac{dL_{(i,j),j}}{L_{(i,j)}} = \eta_{(i,j),j} \alpha_{j \to (i,j)} Q_j \frac{dM_e}{M_j} \frac{M_i}{M_i+M_j}.
\end{gather}
Then similar to our other approach we can use these two equations to find the total fractional change in angular momentum and thus the change in separation of the binary $(i,j)$. Doing so, while again taking $M_1$ to be the donor, and setting $\eta_{(i,j),i} = \eta_{(i,j,j)} = \eta_{ij}$
we get the following three equations:
\begin{gather}
    \frac{a_{12}(f)}{a_{12}(0)} = \left(\frac{M_1(0)+M_2(0)}{M_1(f)+M_2(f)}\right)^{-1} \left(1 + \frac{M_1^{\mathrm{env}}}{M_1^c}\right)^{2} \exp \left(-2 \eta_{12}  \frac{ M_1^{\mathrm{env}}}{M_1^c+M_2}  \mathcal{F}_{12}\right),\\
    \frac{a_{13}(f)}{a_{13}(0)} = \left(1 + \frac{M_1^{\mathrm{env}}}{M_1^c + M_3}\right)^{-1} \left(1 + \frac{M_1^{\mathrm{env}}}{M_1^c}\right)^{2} \exp \left(-2 \eta_{13} \frac{M_1^{\mathrm{env}} }{M_1^c+M_3} \mathcal{F}_{13}\right),\\
    \frac{a_{23}(f)}{a_{23}(0)} = \exp \left(-2 \eta_{23} \frac{M_1^{\mathrm{env}}}{M_2+M_3} \mathcal{F}_{23}\right),
\end{gather}
where we define
\begin{equation}
    \mathcal{F}_{ij} = \alpha_{i \to (i,j)} Q_i \frac{M_j}{M_i} + \alpha_{j \to (i,j)} Q_j \frac{M_i}{M_j}.
\end{equation}
One can verify that these three equations are mathematically identical to those presented in the main body of the paper, showing how our approach holds regardless of the order in which we choose to partition the CE between the individual stars and binaries.

\section{Additional Figures} {\label{appendix:figures}}
We provide additional two-dimensional projections analogous to those shown in Figure~\ref{fig:sc2input}. These plots span the same range of $A$ and $B$ values and display the ratio of binary component masses along the horizontal axis and the envelope-to-core mass ratio along the vertical axis just as done in Figure~\ref{fig:sc2input}. For each case, we generate slices at five representative values of the third relevant mass ratio, varying from $10^{-2}$ to $10^{2}$. This range provides us with a comprehensive sampling of the parameter space. As in Figure~\ref{fig:sc2input}, we overlay contours indicating shrinkage factors of $10^{-2}$, $10^{-1.5}$, $10^{-1}$, and unity.

\balance
\clearpage          
\onecolumn          
\begin{figure*}  
  \centering
  \includegraphics[width=.88\textwidth]{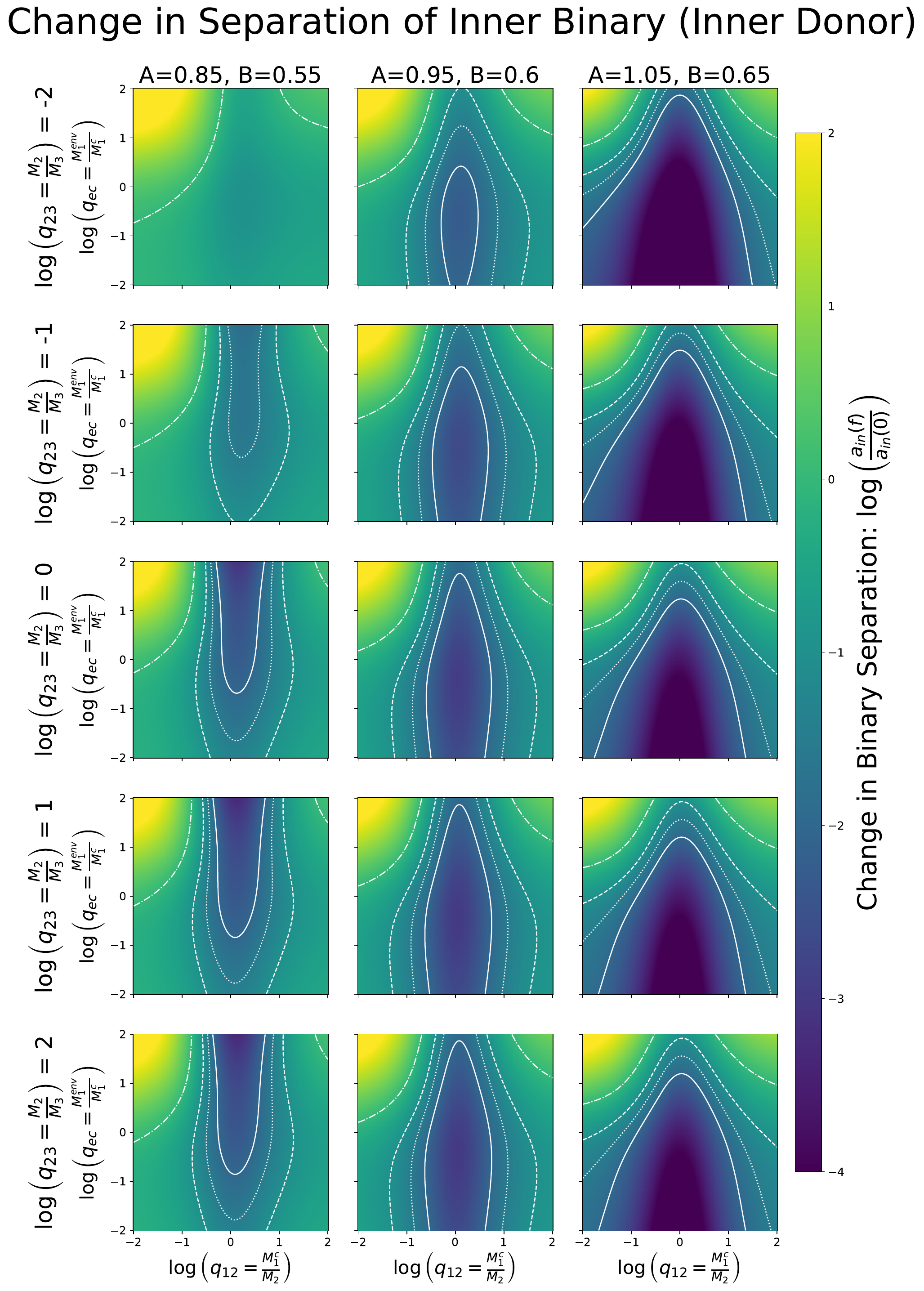}
  \caption{Dependence of change in inner-binary separation when the inner star ($M_1$) serves as the envelope donor. The plots are generated using Equation~\ref{eq:18} and employ three unique fits for the parameter $\eta$ ($(A=0.85 , B=0.55)$, $(A=0.95 , B=0.6)$, and $(A=1.05 , B=0.5)$). We choose to create our plots as functions of the mass ratio between binary components, and between the envelope and core mass as in Figure~\ref{fig:sc2input}, and we reproduce our plots over a reasonable range for the third system parameter $q_{23}$. Contours mirror those of Figure~\ref{fig:sc2input} again with regions within the solid line most likely corresponding to systems which merge shortly after (or during) the CE.}
  \label{fig:ainslices}
\end{figure*}
\begin{figure*}  
  \centering
  \subsection{Non-Hierarchical Systems}
\includegraphics[width=.88\textwidth]{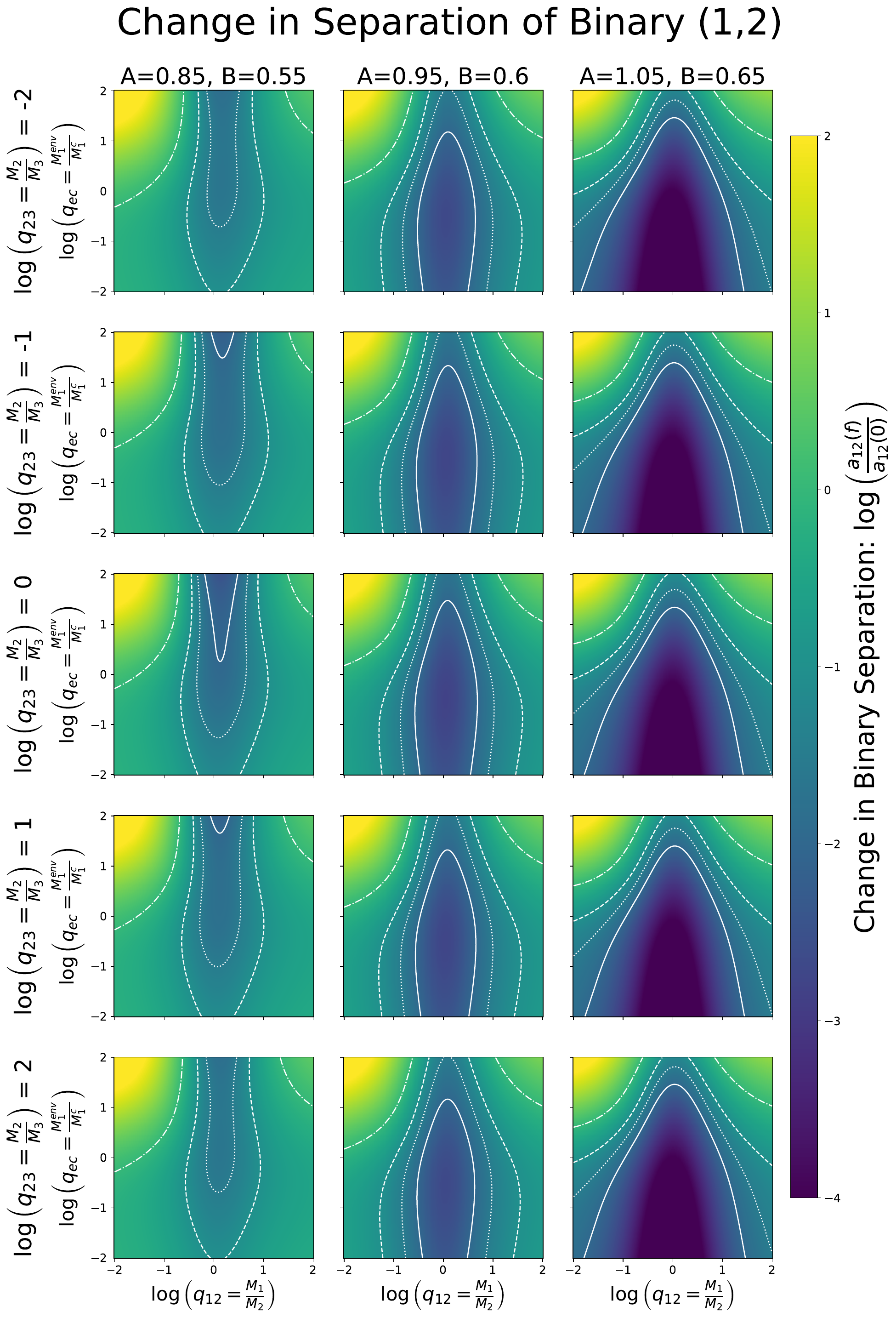}
  \caption{Change in separation for binary (1,2) of non-hierarchical systems as a result of the CE, following Equation~\ref{eq:a12} identical in formatting to Figure~\ref{fig:ainslices}.}
  \label{fig:a12slices}
\end{figure*}
\begin{figure*}  
  \centering
  \includegraphics[width=.88\textwidth]{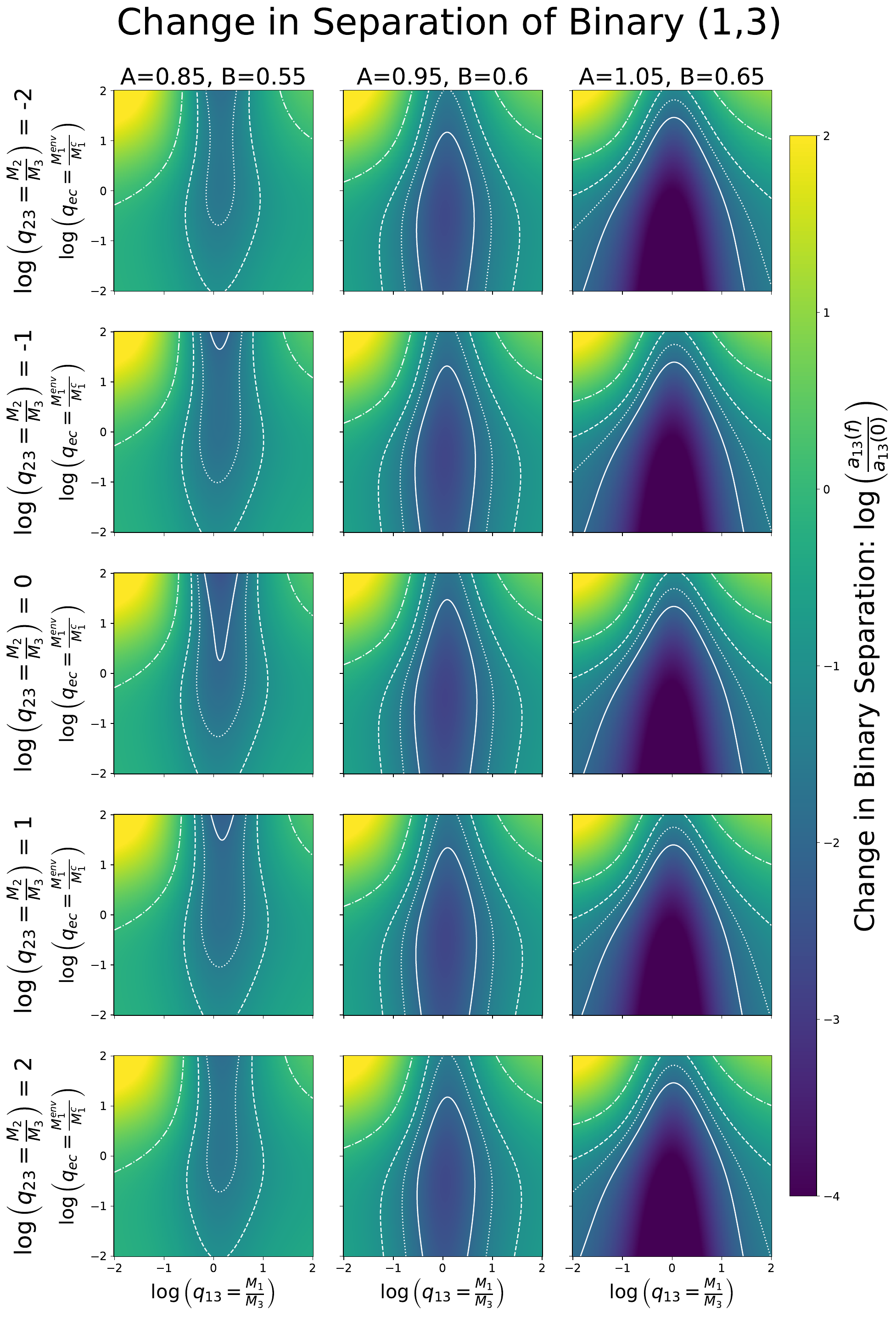}
  \caption{Same as Figure~\ref{fig:a12slices} but for binary (1,3).}
  \label{fig:a13slices}
\end{figure*}
\begin{figure*}  
  \centering
  \includegraphics[width=.88\textwidth]{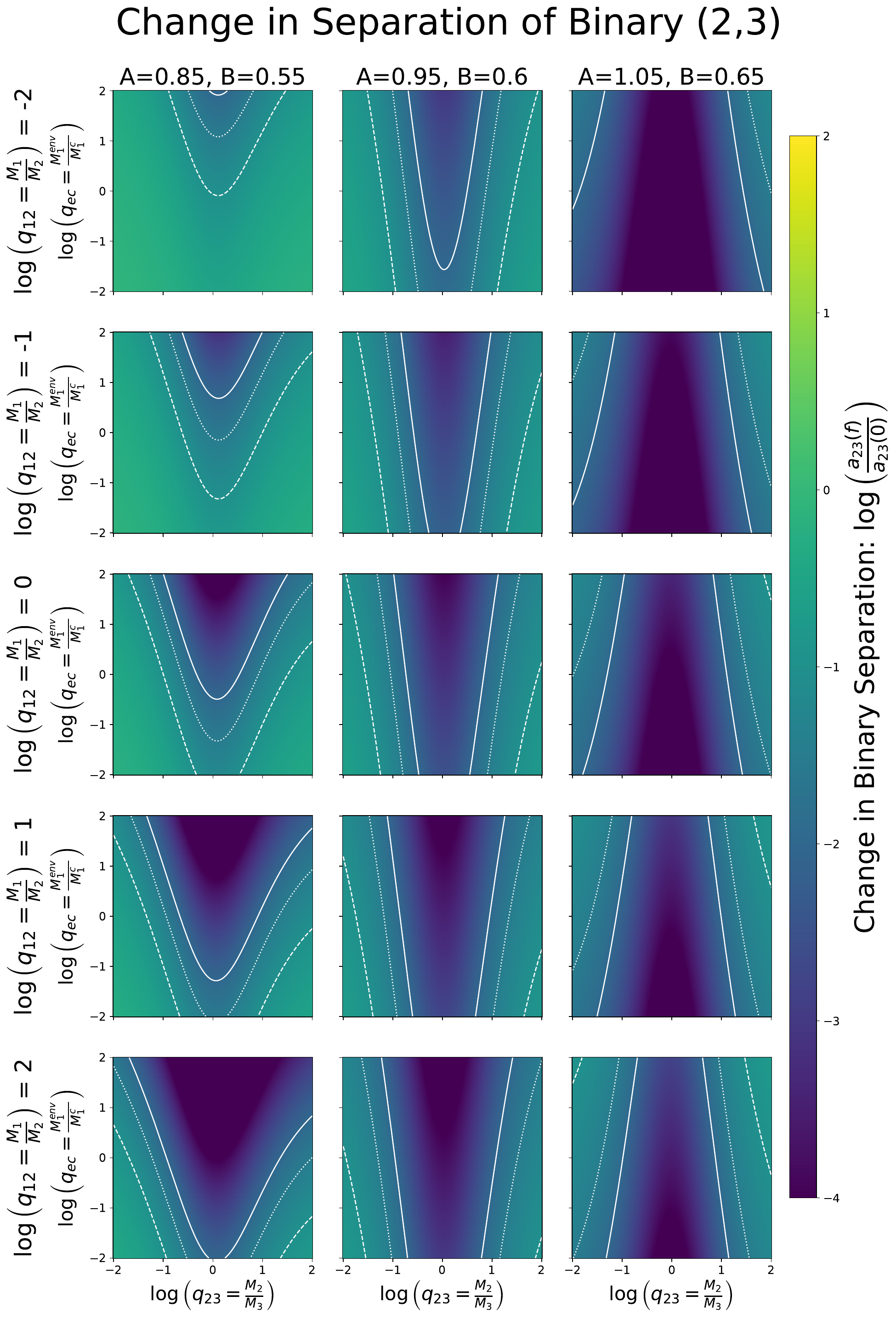}
  \caption{Same as Figure~\ref{fig:a12slices} but for binary (2,3).}
  \label{fig:a23slices}
\end{figure*}

\twocolumn         
\bibliographystyle{mnras}
\bibliography{CE_refs.bib} 


\bsp	
\label{lastpage}
\end{document}